\documentclass[aip,amsmath,amssymb,preprint]{revtex4-1}
\usepackage[colorlinks=true, urlcolor=blue, linkcolor=red]{hyperref}
\usepackage{graphicx}
\usepackage{dcolumn}
\usepackage{bm}
\usepackage{amsthm}
\usepackage{eqnarray,amsmath}
\usepackage[utf8]{inputenc}
\usepackage[T1]{fontenc}
\usepackage{mathptmx}
\usepackage{float}
\usepackage{enumitem}
\usepackage{dsfont}
\usepackage{caption}
\usepackage{subcaption}
\usepackage{appendix}
\usepackage{multirow}

\newcommand{\av}{\textrm{{av}}}

\newcommand{\dr}{\textrm{{dr}}}

\newcommand{\TT}{\textrm{\tiny{T}}}
\newcommand{\BB}{\textrm{\tiny{B}}}
\newcommand{\cTn}{\textit{cTn}}
\newcommand{\acTn}{$\overline{\textit{cTn}}$}
\begin{document}

\preprint{AIP/123-QED}
\title[]{The Role of Biomarkers on Haemodynamics in Atherosclerotic Artery}
\author{Ruchira Ray}
\affiliation{Department of Mathematics, University of North Bengal, Raja Rammohunpur, Darjeeling-734013, West Bengal, India}
\author{Bibaswan Dey}
\email{bibaswandey@nbu.ac.in}
\affiliation{Department of Mathematics, University of North Bengal, Raja Rammohunpur, Darjeeling-734013, West Bengal, India}
\begin{abstract}
Atherosclerosis, a chronic inflammatory cardiovascular disease, leads to arterial constriction caused by the accumulation of lipids, cholesterol, and various substances within artery walls. Such plaque can rupture, resulting in a blood clot that obstructs major arteries and may initiate myocardial infarction, ischemic stroke, etc. Atherosclerotic plaque formation begins with the accumulation of foam cells and macrophages within the intima layer of the arterial wall. At the latter stage, the smooth muscle cells migrated from deeper artery wall layers, contributing to the fibrous cap formation and plaque stabilizing. A developed plaque gradually enters the lumen and narrows down the lumen to impede blood flow. We introduce a two-phase and macroscopic model to investigate the progression of plaque growth in its advanced stage and analyze the minimum gap (Lumen Clearance) within an atherosclerotic artery so that blood cells can pass through. Cardiac troponin, a high specificity and sensitivity biomarker, facilitates early detection of elevated myocardial infarction, Ischemic stroke, etc. risks. This study aims to establish a relationship between the troponin concentration in atherosclerotic arteries and their internal clearance, which could significantly improve our understanding of disease progression. Our observations show that the plaque undergoes rapid evolution in its initial stages, gradually slowing down over time to reach a steady state. At the same time, the lumen clearance exhibits an opposite behavior, decreasing slowly over time. Our study finds a positive correlation between plaque depth and troponin concentration in the blood and a negative relationship between troponin concentrations and lumen clearance in atherosclerotic arteries.
\end{abstract}
\maketitle
\section{Introduction}
\noindent
Cardiovascular diseases play a significant role in the mortality of individuals worldwide, establishing them as a prominent contributor to human death on a global scale. \cite{roth2018global}. The World Health Organisation (WHO) reports that around 17.9 million individuals pass away each year due to CVDs. Atherosclerosis is a cardiovascular disease in which a plaque made by fat and lipids forms within the large and medium-sized conduit arteries with a bifurcated structure, mainly where the disturbed flow and low values of endothelial shear stress (ESS) occur \cite{chatzizisis2007role,khan2023effect}. Individuals who experience conditions such as hypertension, diabetes, hyperlipidemia, chronic infections, and similar health issues have a higher probability of developing atherosclerosis, as stated by \citet{cunningham2005role}. This disease starts with the accumulation of large lipid-laden foam cells derived from monocytes, which engulf oxidized low-density lipoprotein (ox-LDL) present in the intima, leading to atherosclerotic plaque formation \cite{avgerinos2019mathematical}. As time passes, the plaque undergoes a gradual and progressive enlargement, increasing bulkiness. This growth can lead to the extension of the plaque into the lumen, subsequently obstructing blood flow and posing a potential risk for myocardial infarction or stroke. In order to understand the progressive hindrance of blood flow caused by plaque, it is crucial to have a comprehensive understanding of the fundamental process involved in plaque formation.

\begin{figure}[htp]
\centering
\includegraphics[width=0.85\linewidth]{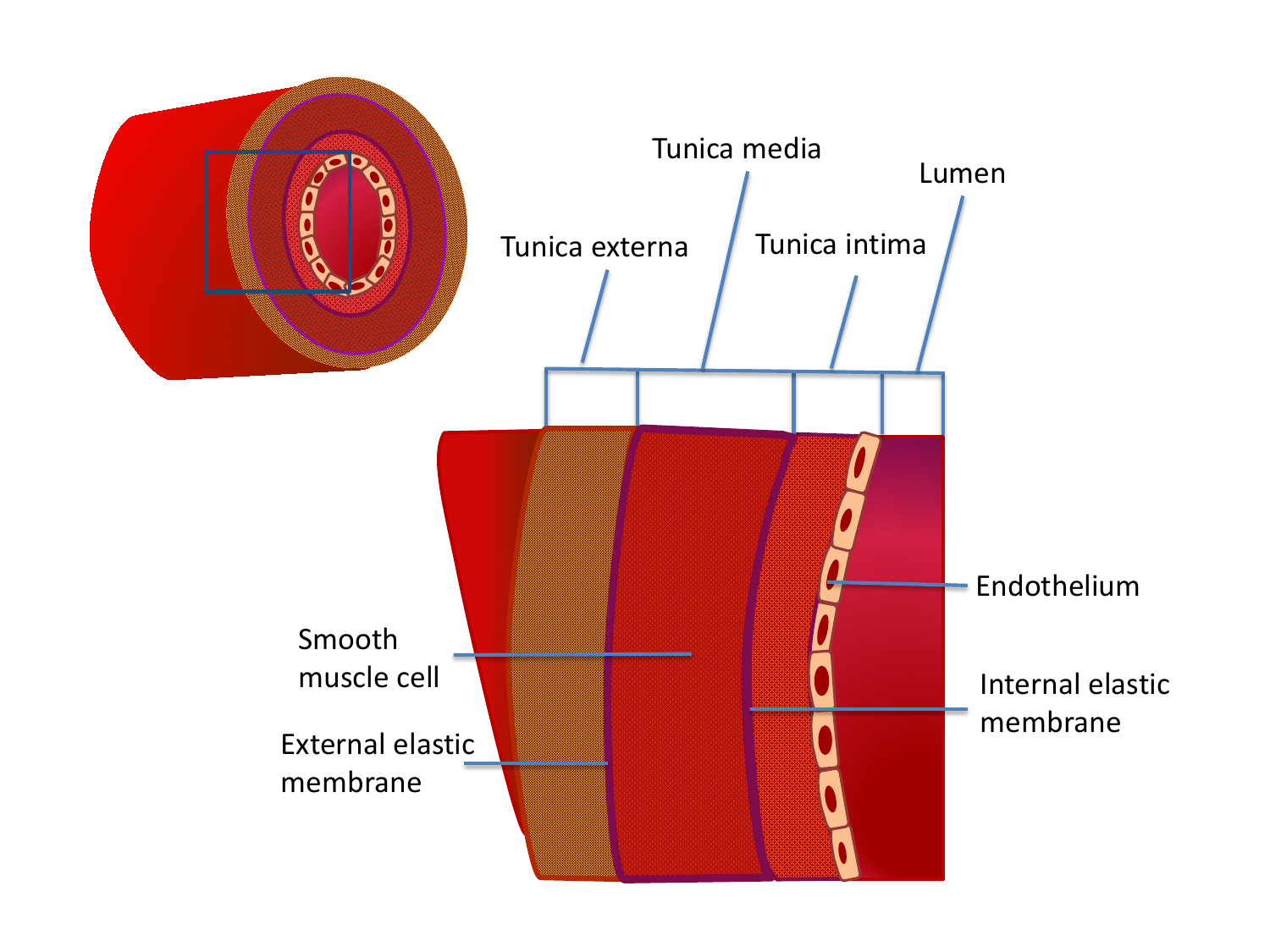}
\caption{{In the cross-section of a healthy artery, the artery wall is composed of distinct cell layers. The innermost layer, the endothelium, is a thin sheet of endothelial cells. Below the endothelium lies the intima layer, followed by the media layer, which is separated from the intima by the internal elastic lamina (IEL). The outermost layer is referred to as the tunica externa.}}
\label{fig:artery}
\end{figure}
\noindent
Atherosclerotic plaque develops within the narrow intimal layer of the artery wall, located between the endothelium and the media layer. Figure \ref{fig:artery} illustrates the layered structure of a healthy artery wall. Endothelial cells (ECs) change their shape due to different levels of wall shear stress, precisely, a high wall shear stress causes elongation of ECs, while low or oscillating wall shear stress results in round or polygonal shapes without a specific alignment pattern \cite{shaaban2000wall}. These observed changes in ECs morphology, particularly in the presence of low endothelial shear stress (ESS), may contribute to widening gaps or spaces between ECs \cite{chatzizisis2007role}. Through these wide gaps or spaces between ECs, low-density lipoproteins (LDL) from the bloodstream can infiltrate the artery wall to enter the intima. Within the intima, LDL molecules are modified by free oxygen radicals and transform into oxidized LDL (oxLDL) \cite{cobbold2002lipoprotein,lusis2000atherosclerosis}. During the oxidation process, oxidative stress causes endothelial cells to express leukocyte adhesion molecules like vascular cell adhesion molecule-1 (VCAM-1) and intracellular adhesion molecules (ICAM-1) \cite{libby2002inflammation,hansson2006immune}. These molecules facilitate the attachment of primarily monocytes, leukocytes, from the bloodstream to the endothelium \cite{lusis2000atherosclerosis}. Additionally, the presence of oxLDLs also triggers endothelial cells (ECs) and smooth muscle cells (SMCs) to release monocyte chemotactic protein-1 (MCP-1) and monocyte colony-stimulating factor (M-CSF) \cite{wu2017new}. In response to the chemoattractant gradients like MCP-1 and other cytokines, monocytes migrate into the tunica intima layer \cite{han2004crp,hansson2006immune}.\\

\begin{figure}[htp]
\centering
\includegraphics[width=\linewidth]{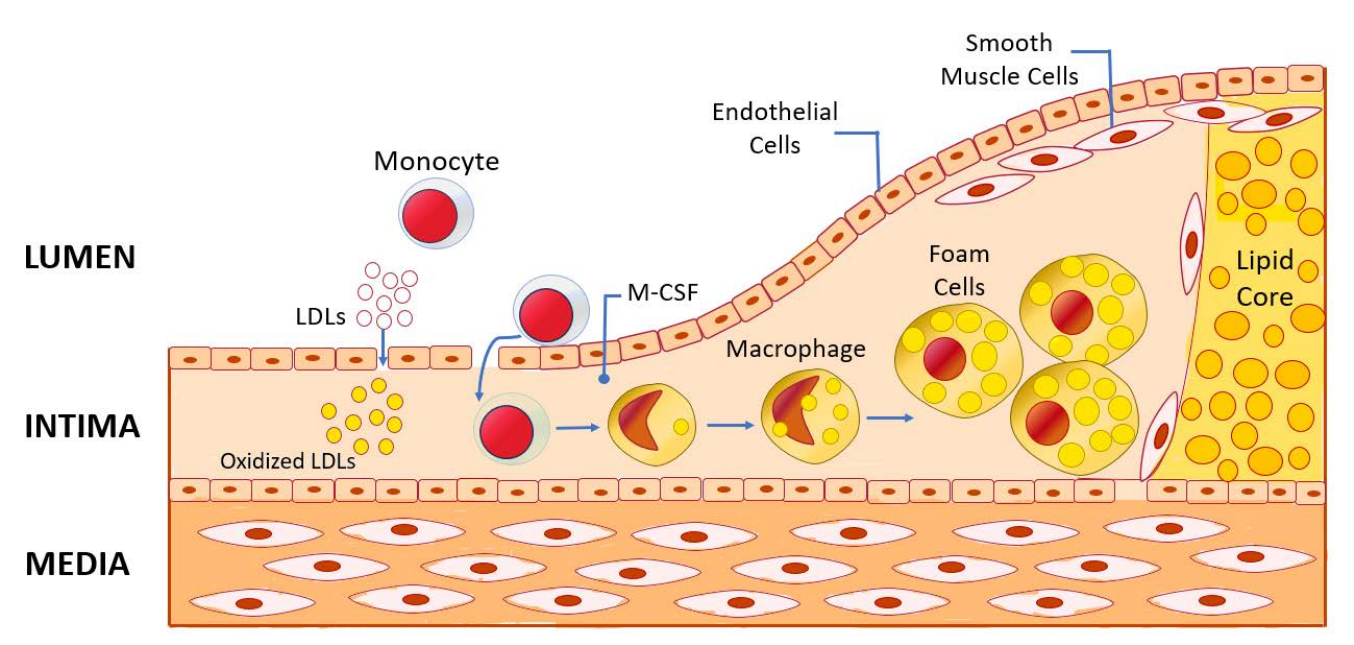}
\caption{Schematic Diagram of the Plaque Formation Process}
\label{fig:plaque1}
\end{figure}
\noindent
The monocytes within the intima differentiate into macrophages after being activated by M-CSF \cite{ross1999atherosclerosis,libby2002inflammation}. These macrophages bind to oxLDL through scavenger receptors (SR) or toll-like receptors (TLR) \cite{avgerinos2019mathematical}. After internalizing oxLDL, the macrophages transform into large lipid-laden foam cells \cite{lusis2000atherosclerosis,gui2012diverse,calvez2009mathematical}. To draw more monocytes into the intima layer, these foam cells release more cytokines and chemical signals \cite{chalmers2017nonlinear,ross1999atherosclerosis}. When these foam cells die inside the plaque, they deposit lipid content and other cellular debris \cite{berliner1995atherosclerosis}. This promotes additional macrophage recruitment and initiates a cycle of chronic inflammation that aids in the development and growth of the plaque. Subsequently, smooth muscle cells (SMCs) located in the media layer migrate toward the endothelium (intima layer) in response to the activation of endothelial cells caused by oxidative stress and the secretion of platelet-derived growth factor (PDGF). In addition,  the lipid-laden macrophages also release PDGF, which further enhances SMC migration \cite{newby1999fibrous}. These SMCs play a role in the production of collagen, which is controlled by the cytokine transforming growth factor-beta (TGF-$\beta$) secreted from endothelial cells, SMCs, and foam cells \cite{hansson2006inflammation}. The smooth muscle cells (SMCs) within the intima and collagen form a cap-like structure called the fibrous cap near the endothelium, stabilising the plaque \cite{newby1999fibrous}. At this stage, the plaque becomes bulkier and may protrude into the lumen, obstructing blood flow in the artery. Figure \ref{fig:plaque1} illustrates the process of plaque formation depicted in a schematic diagram.\\

\noindent
With the increasing size of the plaque, the blood flow rate through the artery decreases, resulting in a reduction in the delivery of oxygen through the blood, which leads to an oxygen-deprived state of the heart muscle. The reduced supply of oxygen can lead to damage in the heart muscle, which is indicated by the release of troponin (troponin I or troponin T) into the bloodstream. Elevated cardiac troponin levels can be a biomarker for detecting atherosclerosis \cite{babuin2005troponin,westermann2017high,anaya2013evolving}. Higher concentrations of troponin indicate an increased risk of heart attack, stroke, and even death. Similarly, C-reactive protein (CRP) is another biomarker helpful in assessing the risk of cardiac risks \cite{adams2004new}. These biomarkers play a crucial role in preventing atherosclerosis-related cardiac complications.\\

\noindent
Beside experimental investigations \cite{young1973flow}, several theoretical studies \cite{chakravarty1989effects,chakravarty1992dynamic,mandal2005unsteady,srivastava2010two,nadeem2011power,roy2017modelling} examine the impact of
stenosis (due to the atherosclerotic plaque developed) in arteries, i.e., the impact of arterial narrowing on blood flow characteristics through the deployment of analytical or numerical tools. \citet{siddiqui2009mathematical} and \citet{sankar2009mathematical} respectively developed mathematical models to describe the blood flow in a stenosed artery, considering two different non-Newtonian fluid models, namely the Casson fluid and the Herschel-Bulkley fluid model. The impact of pulsatile nature of arterial blood flow in a stenosed artery either in absence \cite{mandal2007effect,shaw2010effect} or presence of external field \cite{ponalagusamy2013blood,bunonyo2018modeling} (such as magnetic field, a heat source, etc.). In some of these studies, the arterial wall is modelled as an elastic cylindrical tube. \citet{das2021solute} introduced a mathematical study to investigate solute dispersion in a stenotic tube with a permeable wall, where the blood rheology was modelled using the Casson model and the solute is administered at an evenly distributed point source across the cross-section. A segment of the arterial wall has an axisymmetric cosine-shaped stenosis that narrows the lumen. Such a stenosis geometry was proposed earlier by \citet{andersson2000effects} and \citet{yakhot2005modeling}. Two recent studies, conducted by \citet{dhange2022mathematical} and \citet{hussain2022mathematical,hussain2023numerical}, have analyzed the blood flow in a stenosed artery with a cosine-shaped stenosis geometry. While the former considers the blood as a Casson-type non-Newtonian fluid, the latter deliberates it Newtonian (as blood behaves like a Newtonian fluid in larger cavities and arteries) \citep{hussain2022mathematical,hussain2023numerical}. With an increase in stenosis elevations, both the blood flow rate and wall shearing stress decrease. Moreover according to \citet{hussain2023numerical}, the shape of the stenosis has a crucial impact on the blood flow velocity, total pressure reduction, etc. Blood velocity decreases over time when the walls become narrower and grows along the symmetrical axis. None of the above studies takes into account the progression of stenosis over time and the consequent impact upon the restriction of blood cell movement. In this context, one may refer to the studies conducted by \citet{damiano1998effect} and \citet{damiano1996axisymmetric,damiano2004estimation} on the rheology of blood/capillary systems. These studies show that capillaries with a thicker endothelial cell-glycocalyx layer (EGL) tend to have higher resistance and lower capillary tube hematocrit (volume-fraction of red cells in the blood) than those with a thinner EGL. At high shear rates, whole blood behaves as a Newtonian fluid with an apparent viscosity that depends on hematocrit. The blood viscosity experiences a significant surge as the diameter of the vessels decreases.\\

\noindent
Next we would like to focus on remodelling the growth of the atherosclerotic plaque within stenosed arteries. A growing plaque contains various components: inflammatory tissues (monocytes, macrophages, foam cells, and smooth muscle cells (SMCs)), interstitial fluids, extracellular matrix (ECM) proteins, cytokines, lipids, oxidised LDL (oxLDL), other generic tissues, etc. Therefore, it exhibits a multiphase nature. In this regard, the multiphase mixture theory \cite{byrne2003two,rajagopal2007hierarchy,hubbard2013multiphase,dey2016hydrodynamics,kumar2020elastohydrodynamics,dey2021mathematical,pramanik2023two,kumar2018nutrient} provides the framework for mathematical modelling, especially when dealing with materials that exhibit a multiphase nature, indicating that they comprise multiple constituents. \citet{watson2018two} utilize the mixture theory approach to model the early development of the fibrous cap in the atherosclerotic artery. In another study, \citet{watson2020multiphase} introduce a model on the formation of atherosclerotic cap assuming that the plaque tissue is a mixture of three phases: SMC phase, extracellular matrix (ECM) protein (made by the collagen-rich fibrous tissue) and a generic tissue phase (containing the remaining constituents of plaque, i.e., lipids, foam cells etc.). In this investigation, plaque formation in the atherosclerotic artery is modelled using the multiphase approach. In early atherosclerosis, monocytes, macrophage foam cells, etc. are the principal inflammatory components of the plaque \cite{chalmers2015bifurcation,ahmed2023macrophage}. On the contrary, the migration of SMCs from the media layer of the blood vessel wall to the plaque region under the action of chemoattractant called Platelet-derived growth factor (PDGF) contributes significantly towards the hardening of plaque \cite{hao2014ldl,watson2018two} alongside the synthesis of collagen-fibre. In such case, it becomes the secondary plaque with the formation of fibrous cap. In this study, we are basically interested with a fully grown plaque that can prevent the motion of various blood cells.\\

\noindent
Numerous studies have investigated plaque formation in atherosclerotic arteries, examining the effects of stenosis on blood flow characteristics within arteries. However, literature regarding the influence of growing atherosclerotic plaque on the evolution of biomarkers in blood is scarce. Various recent literary works concentrate on the narrowed internal geometry of arteries undergoing atherosclerosis and its impact on blood flow characteristics. For example, \citet{pandey2022effect} develop a model with different blood viscosity prototypes along with different Reynolds numbers corresponding to the blood flow to investigate hemodynamic outcomes of the diseased multi-stenosed Left Coronary Artery. \citet{owais2023effect} examine the influence of a bend on vortex formation and development within a stenosed artery in 3D geometry under the pulsatile motion of blood. \citet{sakthivel2022three} formulated a three-dimensional characteristic-based off-lattice Boltzmann method in generalized cylindrical curvilinear coordinates to simulate blood flow through an irregularly stenosed artery. In this context, authors think that one should focus on models correlating the temporal progression of stenosis (plaque) in atherosclerotic arteries with the release of biomarkers like cardiac troponin into the bloodstream (due to plaque growth). Consequently, we develop a mathematical model to take into account the progression of stenosis over time, the consequent impact on the blood cell movement through the minimum gap within a narrow artery, and aims to establish a connection between maximum plaque growth and the increment in cardiac troponin levels in the bloodstream. The growing plaque can be assumed as a two-phase material consisting of inflammatory and non-inflammatory tissues. The governing equation that describes the lumenal flow blood plasma can be considered from the Navier-Stokes equation under the assumption of Stokes hypothesis, incompressibility, and insignificant inertial effect. Therefore, the blood plasma is considered as a linear viscous fluid since blood behaves like a Newtonian fluid in larger cavities and arteries \cite{damiano1996axisymmetric,damiano1998effect,wu2014numerical,wu2017transport,hussain2022mathematical,hussain2023numerical}. In contrast, blood cells move in a plug-like fashion through a tubular region along the cylindrical axis of blood vessels. The overall hemodynamic model can be formulated in axisymmetric cylindrical coordinates. Increased levels of troponin are a reliable indicator of a higher risk for heart attacks and strokes. Establishing a connection between troponin levels in an atherosclerotic artery and its clearance rate can help identify the early stages of heart attacks and strokes related to atherosclerosis. This study aims to introduce a simple method for preventing such serious complications through basic blood tests.
\section{Mathematical Formulation}
\noindent
In general, atherosclerosis is an inflammatory process involving the active participation of inflammatory cells or immune cells from the blood. Atherosclerotic plaque is comprised of a combination of various constituents; however, in this particular model, the plaque is treated explicitly as a mixture consisting of two distinct constituent phases: One contains monocytes, macrophages, foam cells, Smooth muscle cells (SMCs), etc., while the other consists of generic tissue materials, e.g., lipids, oxLDL molecules, cytokines, extracellular matrix (ECM) proteins, interstitial fluid and other components. The first constituent phase is designated as the inflammatory tissue phase, as it directly participates in plaque growth, while the second is the non-inflammatory cell phase. However, each of the above-stated components may have some role to play in plaque stabilization and hardening. In particular, in this study, we do not explicitly consider any ECM remodelling (quantitative and qualitative changes in the ECM, mediated by specific enzymes responsible for ECM degradation, such as metalloproteinases). However, ECM degradation may contribute to the production of generic tissue materials mentioned in this study.\\

\noindent
In this study, we assume that the geometry of developed plaque (stenosis) follows axisymmetry and cosine shape within the lumen of an arterial segment $[-L,~L]$ over time. In addition, the boundary of the plaque region exhibits a specific geometric shape, which is characterized by \cite{das2021solute}
\begin{equation}\label{eqn0}
B(z,t)=\begin{cases}
d_0-\frac{\xi(t)}{2}\left(1+\cos\left(\frac{\pi z}{L}\right)\right) \quad & -L\leq z \leq L, \\
d_0 \quad &\text{otherwise},  \\
\end{cases}
\end{equation}
\begin{figure}[htp]
\centering
\includegraphics[width=\linewidth]{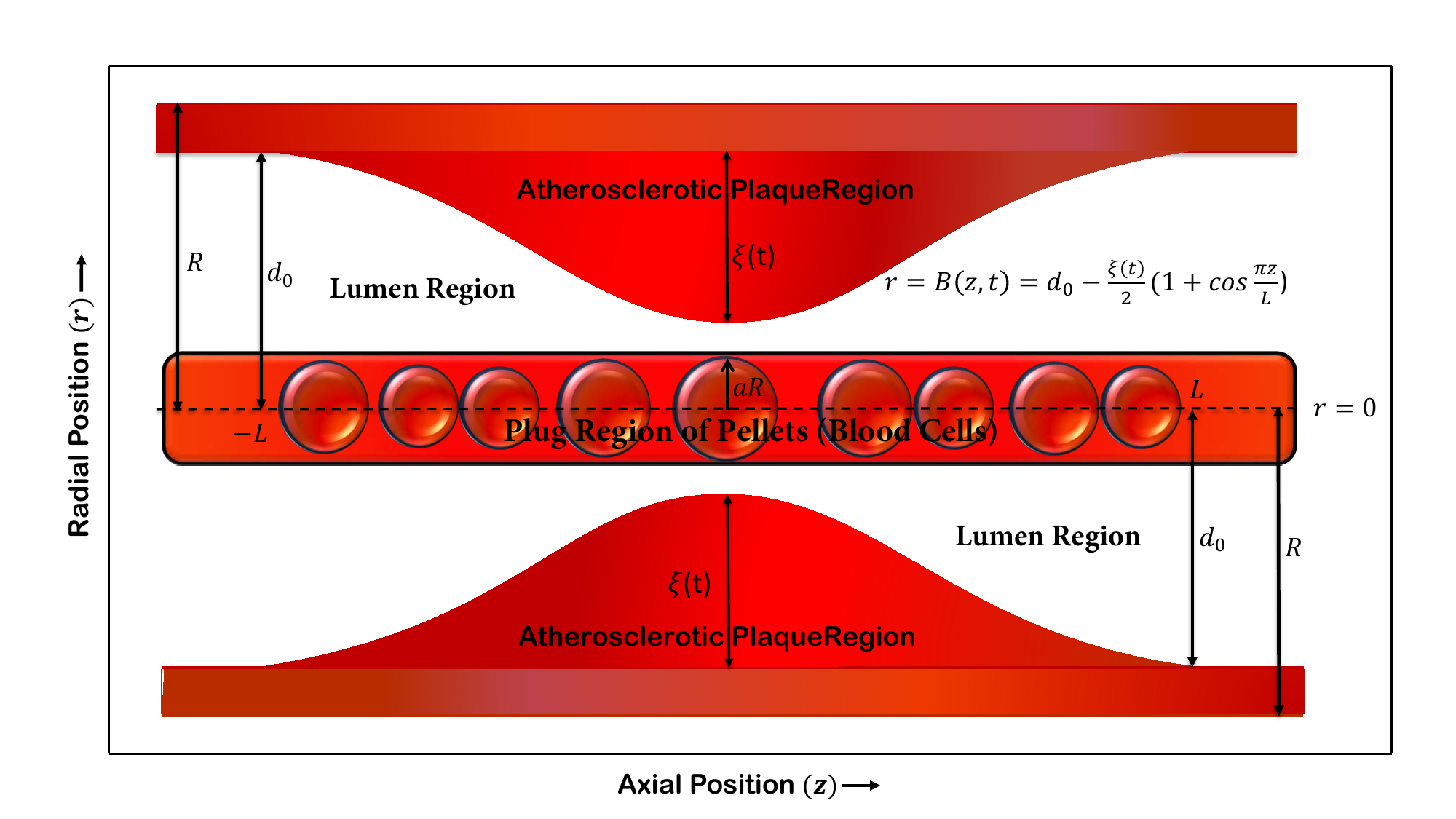}
\caption{{It is a schematic representation of an artery with a cosine shaped stenosis, along with the flow of axisymmetric pellets in the form of a cylindrical plug region having a diameter strictly less than the minimum diameter of the free lumen. The radius of the cylindrical plug region is given by $aR$ where $a<1$ (regulates the number density of pellets).}}
\label{fig:plaque2}
\end{figure}
where $L$ is the half-length, $d_0$ is the radius of the lumen of a healthy artery (without stenosis), and $\xi(t)$ represents the maximum plaque (stenosis) depth (\textit{MPD}) that varies over time, indicating a progressive behaviour of atherosclerotic plaque as time elapses. $r=B(z,t)$ is the interface between lumen and plaque, which is also the effective radius of the lumen that depends on the axial coordinate. Therefore, where $r=B(0,~t)$ is the minimum radius of the lumen available. In other words, $2\left|d_0-\xi(t)\right|$ is the minimum available diameter of the arterial lumen such that the red blood cells and other blood cells can pass through the minimum gap. The blood cells (refer as pellets) with a diameter smaller than the minimum gap inside the free lumen can traverse through. In this study, the lumenal clearance (\textit{CL}) is described by $\left|d_{0}-aR-\xi(t)\right|$ which is the remaining available spaces after the safe passage of the pellets in the form of plug motion along the axial direction. $aR$ represents plug region radius (\textit{PRR}) such that $a$ (with $a<1$) regulates the number density and sizes of the pellets within the plug region. \textit{CL} decreases gradually over time and $\underset{{t\geq 0}}{\min}\left|d_{0}-aR-\xi(t)\right|$ represents the adequate lumenal clearance (\textit{CL}). Figure \ref{fig:plaque2} depicts the schematic representation illustrating the geometry of the problem.
\subsection{Mixture theory for an atherosclerotic plaque}
\noindent
We define an index set $\textit{C}$ as a finite set that includes the constituents of the plaque region. Under the assumption that the atherosclerotic plaque is composed of inflammatory ($\alpha$) and non-inflammatory ($\beta$) tissues, we can express $\textit{C}=\{\alpha,\beta\}$. In this study for each $(i \in \textit{C})$, we assume $\mathbf{v}^i$ as the velocity vector and $\phi^i$ as the volume fraction of $i$-th phase.
\subsubsection{Mass Balance Equations}
\noindent
Under the assumption of constant true density of both the inflammatory and non-inflammatory tissue phases, the mass balance equations for each phase can be written in the following form
\begin{equation}\label{equ1}
\frac{\partial \phi^i}{\partial t}+\overset{\circ}{\boldsymbol{\nabla}}\cdot\left(\phi^i \boldsymbol{v}^i\right)={G}^i,~~~~~(i \in \textit{C}).
\end{equation}
Note that $\overset{\circ}{\boldsymbol{\nabla}}\equiv(\frac{\partial}{\partial r}+\frac{1}{r},0,\frac{\partial}{\partial z})$ is the axisymmetry form of the gradient operator in cylindrical coordinates, along with $\mathbf{v}^i=(v^{i}_{r},0,v^{i}_{z})$ and $\phi^i=\phi^i(r,z,t)$, such that $\theta$-variation is neglected. Here, ${G}^i$ represents the net source term associated with the decay and accumulation of the $i$-th phase. ${G}^i$ is associated with the inflammatory process where excess LDL molecules from the bloodstream accumulate inside the damaged endothelium. Consequently, monocytes from the bloodstream respond to such inflammatory changes and enter the damaged endothelium. Subsequently, they differentiate into macrophages, which consume oxLDLs produced through the oxidation of LDL molecules by the free oxygen radical. One may consider both macrophages and foam cells as the inflammatory tissue phase, while oxLDL is part of non-inflammatory tissue. Furthermore, the death of inflammatory foam cells leads to the accumulation of cellular debris, which becomes part of the non-inflammatory phase. That implies that the demise or depletion of one phase contributes to the birth or emergence of the other phase. Each ${G}^i$ depends on the concentrations of oxLDL, cytokines like MCP-1 (monocyte chemotactic protein 1), M-CSF (macrophage colony-stimulating factors), PDGF (platelet-derived growth factor), (they are respectively denoted as $C_{\tiny{oxLDL}}$, $C_{\tiny{MCP-1}}$, $C_{\tiny{M-CSF}}$, and $C_{\tiny{PDGF}}$) etc. The entire plaque development process is considered a closed system in our modelling. Consequently, one can write
\begin{equation}\label{equ2}
\sum_{i \in \textit{C}} {G^i}=0.
\end{equation}
Further, we assume that there is no void present within the plaque, thus satisfying the following condition
\begin{equation}\label{equ3}
\sum_{i \in \textit{C}} \phi^i = 1.
\end{equation}
\subsubsection{Momentum Balance Equations}
\noindent
In the absence of external body forces and negligible inertia, the momentum balance equations can be simplified to express a balance between intraphase and interphase forces for each inflammatory and non-inflammatory phase:
\begin{equation}\label{equ5}
\overset{\circ}{\boldsymbol{\nabla}}\cdot\left(\phi^i \mathbb{T}^i\right)+\boldsymbol{I}^{i}=0,~~~~ \left(i\in \textit{C}\right),
\end{equation}
where $\mathbb{T}^i$ is the Cauchy stress tensor for the $i$-{th} phase and the term $\boldsymbol{I}^{i}$ represents the interphase interaction force exerted on the $i$-th phase by the other phase(s), where we note that the other phase exerts an equal and opposite force on the $i$-{th} phase. Thus we write
\begin{equation}\label{equ6}
\boldsymbol{I}^{i}=-\boldsymbol{I}^{j},~~~~i,j\in \textit{C},i\neq j.
\end{equation}
In addition, we assume that the interaction force on the $i$-th phase is proportional to the velocity difference between the phases, as described by the following relation \cite{ambrosi2002closure,byrne2003modelling,dey2016hydrodynamics,kumar2018nutrient,dey2021mathematical,pramanik2023two}:
\begin{equation}\label{equ7}
\boldsymbol{I}^{i}= P \overset{\circ}{\boldsymbol{\nabla}} \phi^i+\sum^{j\neq i}_{j \in \textit{C}}\mathbb{K}_{ij}\left(\boldsymbol{v}^j-\boldsymbol{v}^i\right),
\end{equation}
\noindent
where $P$ is the hydrodynamic pressure of the interstitial fluid and $\mathbb{K}_{ij}$ refers to the local interaction in the form of a matrix between the $i$-th and $j$-th constituent. Therefore, $\mathbb{K}_{\alpha\beta}$ is the coefficient matrix corresponding to the momentum transfer between the inflammatory and non-inflammatory tissues in the absence of Fickian diffusion of the individual tissue constituent \cite{ambrosi2002closure,byrne2003modelling,dey2016hydrodynamics,pramanik2023two}. However, the anisotropy nature of $\mathbb{K}_{\alpha\beta}$ is beyond the scope of this study. Hence, $\mathbb{K}_{\alpha\beta}=K_{\alpha\beta}=K_{\beta\alpha}=K$ is the corresponding constant isotropic value.
\subsubsection{Stress Components}
\noindent
Assuming both phases to be isotropic, the arterial plaque is modeled as a composition of an inflammatory cell phase with a dynamic viscosity of $\mu^\alpha$ and a non-inflammatory phase. The constitutive relations for the stress tensors $\mathbb{T}^i~~(i=\alpha,\beta)$ on the $i$-th constituent can be expressed as \cite{damiano1996axisymmetric}
\begin{eqnarray}\label{equ8}
\mathbb{T}^\alpha&=&- \left[P + \frac{2}{3} \mu^\alpha\left(\overset{\circ}{\boldsymbol{\nabla}}\cdot\boldsymbol{v}^\alpha\right)+J\left(\Xi(\phi^{\alpha}),C_{\tiny{MCP-1}},C_{\tiny{M-CSF}},C_{\tiny{PDGF}},..\right)\right]\mathbb{I}\\\nonumber
&+&\mu^\alpha\left(\overset{\circ}{\boldsymbol{\nabla}} \boldsymbol{v}^\alpha+\left(\overset{\circ}{\boldsymbol{\nabla}} \boldsymbol{v}^\alpha\right)^T\right),
\end{eqnarray}
\begin{equation}\label{equ9}
\mathbb{T}^\beta=- P \mathbb{I},
\end{equation}
\noindent
where $P$ is the hydrodynamic pressure; $\boldsymbol{v}^\alpha$ is the velocity vector for the inflammatory cell phase; $J$ is a scalar function that contributes to the stress field to include interaction between inflammatory cells $\left(\Xi(\phi^{\alpha})\right)$ and chemotaxis due to the various cytokines. We view the inflammatory cell and non-inflammatory phases as incompressible fluids and assume they can be treated as viscous and inviscid fluids, respectively, on the timescale of interest. In Eq. (\ref{equ8}), we consider the Stokes hypothesis \cite{dey2016hydrodynamics} and neglect the bulk viscosity of the inflammatory cell phase by taking $2\lambda^\alpha+3\mu^\alpha=0$, i.e. $\lambda^\alpha =-({2}/{3})\mu^\alpha$.
\subsubsection{Plasma Motion and Plug Motion of Pellets through Lumen}
\noindent
In this study, we assume that the plasma motion within the lumen exhibits incompressible Newtonian behavior \cite{damiano1996axisymmetric,damiano1998effect,damiano2004estimation}. The momentum balance equation of the fluid in the lumen is described by using the Stoke's equation of motion without the inertia effect as follows:
\begin{equation}\label{equ15}
\mu^l\overset{\circ}{\boldsymbol{\Delta}}\boldsymbol{v}^l=\overset{\circ}{\boldsymbol{\nabla}}P^l,
\end{equation}
where $\overset{\circ}{\boldsymbol{\Delta}}\equiv\left(\frac{1}{r}\frac{\partial}{\partial r}\left(r\frac{\partial}{\partial r}\right)+\frac{\partial^{2}}{\partial z^{2}}\right)$ is the axisymmetry form of Laplacian operator in cylindrical coordinates. Eq. \ref{equ15} is followed by the mass conservation equation incorporating the incompressibility assumption stated as
\begin{equation}\label{equ16}
\overset{\circ}{\boldsymbol{\nabla}}\cdot\boldsymbol{v}^l=0,
\end{equation}
where $P^l=P^{l}(r,z)$ denotes the hydrodynamic pressure inside the lumen, $\boldsymbol{v}^l=\left(v^{l}_{r},0,v^{l}_{z}\right)$ represents the velocity vector, and $\mu^l$ corresponds to the dynamic viscosity of plasma in the lumen. Further, within the plug region the pellets used to move with constant velocity $v^{p}\hat{\mathbf{e}}_{z}$ along the axial direction.
\subsubsection{Boundary Conditions}
\begin{enumerate}[label=(\roman*)]
\item At $r=R$ i.e., the outer most boundary of the artery, we impose the no-slip and no-penetration conditions
\begin{equation}\label{equ20}
\boldsymbol{v}^\alpha=0,~~~~~~\textrm{and}~~~~~~\boldsymbol{v}^\beta=0,
\end{equation}
to signify that the overall growth of the plaque material remain confined within the arterial wall. In subsequent time, without lack of space for growth, the plaque material has to invade the lumen.
\item At $r_{0}=B(z_{0},t)$ for some $z_0\in[-L, L]$ i.e., at some point on the interface between the atherosclerotic plaque and the lumen, we assume the continuity of the lumen velocity with the composite velocity of the plaque material \cite{hou1989boundary,damiano1996axisymmetric}:
\begin{equation}\label{equ21}
\phi^\alpha v^\alpha_z+\phi^\beta v^\beta_z=v^l_z.
\end{equation}
On the other hand at $r=r_0$,
\begin{equation}\label{equ23}
v^l_r=0,
\end{equation}
can be imposed due to the fact that the blood-plasma cannot penetrate the plaque material. Now in this model, following \citet{hou1989boundary,damiano1996axisymmetric} a stress continuity is enforced between the plaque and the clear fluid in the lumen, which is done by equating the stress-traction vector, represented by $\boldsymbol{\hat{n}}.\mathbb{T}^\alpha$, on the interfacial surface associated with the $\alpha$-th phase to the product of the volume fraction of that phase and the stress-traction vector of the blood-plasma in the lumen. The same condition can also be applied to the $\beta$-th phase. Accordingly,
\begin{equation}\label{equ24}
\boldsymbol{\hat{n}.}\mathbb{T}^\alpha=\phi^\alpha \boldsymbol{\hat{n}.}\mathbb{T}^l,
\end{equation}
and
\begin{equation}\label{equ25}
\boldsymbol{\hat{n}.}\mathbb{T}^\beta=\phi^\beta \boldsymbol{\hat{n}.}\mathbb{T}^l.
\end{equation}
\noindent
Addition of above two conditions obtains
\begin{equation}\label{equ26}
\boldsymbol{\hat{n}.}(\mathbb{T}^\alpha+\mathbb{T}^\beta) = \boldsymbol{\hat{n}.}\mathbb{T}^l,
\end{equation}
\noindent
which represents the equality between the stress-traction vector of the entire plaque material and that of the blood-plasma at the interfacial surface.
\item On $r=aR$, we impose the following condition
\begin{equation}\label{equ27}
\boldsymbol{v}^l.\hat{\mathbf{e}}_{z}=v^p,
\end{equation}
\noindent
where $v^p$ is the velocity of the pellet stream. It suggests the fact that the pellets move primarily towards the axial direction ($z$-direction) with the same axial velocity of lumen. Moreover at $r=aR$, the zero normal stress condition due to the lumenal blood-plasma flow can be assumed on the plug motion of pellets
\begin{equation}\label{equ29}
\boldsymbol{\hat{n}.}\mathbb{T}^l.\boldsymbol{\hat{n}}=0,
\end{equation}
which suggests an equilibrium situation resulting in the balance of total forces on the plug motion of pellets. The size of a pellet is controlled by the constant parameter $a$. The plug motion of a particular type of pellet can be studied at any given time.
\item
The pellet velocity $v^{p}$ can be obtained from the prescribed volumetric flow rate $Q$ condition across a section of the total region (plaque+lumen) given for all $z \in [-L,L]$ as
\begin{equation}\label{equ30}
\int_0^{aR} v^p\,r\,dr+\int_{aR}^{B(z,t)} v^l_z\,r\,dr+\int^R_{B(z,t)}(\phi^\alpha v^\alpha_z+\phi^\beta v^\beta_z )\,r\,dr = RQ.
\end{equation}
\end{enumerate}
\subsubsection{Model Simplification}
\noindent
Eq. (\ref{equ1}) is summed up over $i \in \textit{C}$ to obtain
\begin{equation}\label{equ4}
\overset{\circ}{\boldsymbol{\nabla}}\cdot\left(\phi^\alpha \boldsymbol{v}^\alpha+\phi^\beta \boldsymbol{v}^\beta\right)=0,
\end{equation}
where $\boldsymbol{v}^{\av}=\phi^\alpha \boldsymbol{v}^\alpha+\phi^\beta \boldsymbol{v}^\beta$ is the volume average velocity of the atherosclerotic plaque. Note that in contrast to the primitive mass conservation equation (\ref{equ1}), Eq. (\ref{equ4}) represents an incompressible form corresponding to the entire continuum structure of the plaque without a net source or sink. On the other hand, terms like $G^{i}$ and $J$ contain concentrations of resources for plaque formation, e.g., oxLDL and various cytokines mentioned before. For the microscopic modelling of plaque formation, we must deploy the transport reaction equations corresponding to the materials discussed above into the mathematical modelling. However, for a macroscopic modelling like the present study, the explicit consideration of all the microscopic phenomena complicates the overall study. Moreover, it would be impossible to separate the impact of the hydrodynamic pressure from the chemotaxis due to the oxLDL and various cytokines on plaque growth. In a macroscopic study, hydrodynamic pressure predominates over chemotactic phenomena due to the larger time scale compared to a microscopic analysis. Therefore, the above discussion enables us to consider $|P|\gg |J|$ for a macroscopic modelling of plaque growth. \\

\noindent
Substituting the relations (\ref{equ7}), (\ref{equ8})-(\ref{equ9}) into the momentum conservation equations (\ref{equ5}) and utilizing the above mentioned assumption, we obtain
\begin{eqnarray}\label{equ10}
-\phi^\alpha \overset{\circ}{\boldsymbol{\nabla}} P+ \frac{1}{3}\left(\phi^\alpha\mu^\alpha\right)\overset{\circ}{\boldsymbol{\nabla}} \left(\overset{\circ}{\boldsymbol{\nabla}}.\boldsymbol{v}^\alpha\right)&+&\left(\phi^\alpha\mu^\alpha\right)\overset{\circ}{\boldsymbol{\Delta}}\boldsymbol{v}^\alpha +{K}\left(\boldsymbol{v}^\beta-\boldsymbol{v}^\alpha\right)\\\nonumber
&+&\overset{\circ}{\boldsymbol{\nabla}}\left(\mu^\alpha\phi^\alpha\right)\cdot\left[-\frac{2}{3}\left(\overset{\circ}{\boldsymbol{\nabla}}.\boldsymbol{v}^\alpha\right)+\left\{\overset{\circ}{\boldsymbol{\nabla}}\boldsymbol{v}^\alpha+\left(\overset{\circ}{\boldsymbol{\nabla}}\boldsymbol{v}^\alpha\right)^{\tiny{tr}}\right\}\right]=0,
\end{eqnarray}
\begin{equation}\label{equ11}
-\phi^\beta \overset{\circ}{\boldsymbol{\nabla}} P+ K\left(\boldsymbol{v}^\alpha-\boldsymbol{v}^\beta\right)=0.
\end{equation}
The elevated viscosity of the inflammatory phase of the atherosclerotic plaque depends on the volume of the $\alpha$-th phase. Therefore, the product $\phi^\alpha\mu^\alpha$ represents the weighted viscosity of the plaque material. However, one can adopt the assumption $\phi^\alpha \mu^\alpha\approx\phi^\alpha_{*}\mu^\alpha$, where $\phi^\alpha_{*}$ is the constant value of $\phi^\alpha$ when the aspect ratio of the blood vessel is small enough. A change in $\phi^\alpha$ does not significantly impact $\mu^\alpha$ due to its high magnitude. \\

\noindent
Now Eqs. (\ref{equ10}) and (\ref{equ11}) can respectively be expressed in the component form as follows
\begin{subequations}
\begin{eqnarray}\label{equ10a}\nonumber
-\phi^\alpha\frac{\partial P}{\partial r}+\frac{1}{3}\left(\phi^\alpha_{*}\mu^\alpha\right)\left[\frac{\partial}{\partial r}\left\{\frac{1}{r}\frac{\partial}{\partial r}\left(rv^{\alpha}_{r}\right)+\frac{\partial v^{\alpha}_{z}}{\partial z}\right\}\right]&+&\phi^\alpha_{*}\mu^\alpha\left[\frac{1}{r}\frac{\partial}{\partial r}\left(r\frac{\partial v^{\alpha}_{r}}{\partial r}\right)+\frac{\partial^{2} v^{\alpha}_{r}}{\partial z^{2}}-\frac{v^{\alpha}_{r}}{r^{2}}\right]\\
&+&K\left({v}^\beta_{r}-{v}^\alpha_{r}\right)=0,
\end{eqnarray}
\begin{eqnarray}\label{equ10b}\nonumber
-\phi^\alpha\frac{\partial P}{\partial z}+\frac{1}{3}\left(\phi^\alpha_{*}\mu^\alpha\right)\left[\frac{\partial}{\partial z}\left\{\frac{1}{r}\frac{\partial}{\partial r}\left(rv^{\alpha}_{r}\right)+\frac{\partial v^{\alpha}_{z}}{\partial z}\right\}\right]&+&\phi^\alpha_{*}\mu^\alpha\left[\frac{1}{r}\frac{\partial}{\partial r}\left(r\frac{\partial v^{\alpha}_{z}}{\partial r}\right)+\frac{\partial^{2} v^{\alpha}_{z}}{\partial z^{2}}\right]\\
&+&K\left({v}^\beta_{z}-{v}^\alpha_{z}\right)=0,
\end{eqnarray}
\end{subequations}
and
\begin{subequations}
\begin{equation}\label{equ11a}
-\phi^\beta\frac{\partial P}{\partial r}+K\left({v}^\alpha_{r}-{v}^\beta_{r}\right)=0,
\end{equation}
\begin{equation}\label{equ11b}
-\phi^\beta\frac{\partial P}{\partial z}+K\left({v}^\alpha_{z}-{v}^\beta_{z}\right)=0.
\end{equation}
\end{subequations}
The velocities of the plasma within the lumen, in both the $r$ and $z$-directions, can be expressed in terms of the stream function $\psi$ as follows:
\begin{equation}\label{equ17}
v_r^l=-\frac{1}{r}\frac{\partial \psi}{\partial z}~~\text{and}~~v_z^l=\frac{1}{r}\frac{\partial \psi}{\partial r}.
\end{equation}
Substituting these expressions into equation (\ref{equ15}), we derive the following relationship
\begin{equation}\label{equ18}
E^4\psi\left(r,z\right)=0,
\end{equation}
where
$$E^2\equiv r\frac{\partial}{\partial r}\left(\frac{1}{r}\frac{\partial }{\partial r}\right)+\frac{\partial^2 }{\partial z^2}.$$
\subsubsection{Further Simplified and Dimension-free Model}
\noindent
In order to make the Eqs. (\ref{equ10a})-(\ref{equ11b}), and Eq. (\ref{equ18}) dimensionless, we introduce the following set of dimensionless variables:
$\hat{r}=r/R$, $\hat{z}=z/L$, $\hat{d}_{0}=d_{0}/L$, $\hat{v_r}=v_r/\left(Q/L\right)$, $\hat{v_z}=v_z/\left(Q/R\right)$, $\hat{P}=P/\left(K Q L/R \right)$, $\hat{\psi}=\psi/Q,$ and $\hat{\xi}=\xi/R $ into them to obtain the following sets of dimensionless governing equations expressed as follows (omitting the \emph{hat} symbol)
\begin{subequations}\label{32-37}
\begin{equation}\label{equ32}
\left(\frac{1}{r}+\frac{\partial}{\partial r}\right)\left(\phi^\alpha v^\alpha_r+\phi^\beta v^\beta_r\right)+\frac{\partial}{\partial z}\left(\phi^\alpha v^\alpha_z+\phi^\beta v^\beta_z\right)=0,
\end{equation}
\begin{equation}\label{equ33}
-\phi^\alpha \frac{\partial P}{\partial r}+ \frac{\gamma^2 \delta^2}{3}\left[\frac{\partial}{\partial r}\left\{\frac{1}{r}\frac{\partial}{\partial r}(rv_r^\alpha)+\frac{\partial v_z^\alpha}{\partial z}\right\}\right]+ \gamma^2 \delta^2 \left[\frac{1}{r}\frac{\partial}{\partial r}\left(r \frac{\partial v_r^\alpha}{\partial r}\right)+\delta^2 \frac{\partial ^2 v_r^\alpha}{\partial z^2}-\frac{v_r^\alpha}{r^2}\right]- \delta^2\left( v_r^\alpha-v_r^\beta\right)=0,
\end{equation}
\begin{equation}\label{equ34}
-\phi^\alpha \frac{\partial P}{\partial z}+ \frac{\gamma^2 \delta^2}{3}\left[\frac{\partial}{\partial z}\left\{\frac{1}{r}\frac{\partial}{\partial r}(rv_r^\alpha)+\frac{\partial v_z^\alpha}{\partial z}\right\}\right]+ \gamma^2 \left[ \frac{1}{r}\frac{\partial}{\partial r}\left(r \frac{\partial v_z^\alpha}{\partial r}\right)+\delta^2 \frac{\partial^2 v_z^\alpha}{\partial z^2}\right]- \left(v^\alpha_z-v^\beta_z\right)=0,
\end{equation}
\begin{equation}\label{equ35}
-\phi^\beta \frac{\partial P}{\partial r}+ \delta^2 \left(v^\alpha_r-v^\beta_r\right)=0,
\end{equation}
\begin{equation}\label{equ36}
-\phi^\beta \frac{\partial P}{\partial z}+ \left(v^{\alpha}_{z}-v^{\beta}_{z}\right)=0,
\end{equation}
and
\begin{equation}\label{equ37}
r\frac{\partial}{\partial r}\left(\frac{1}{r}\frac{\partial}{\partial r}\left(r\frac{\partial}{\partial r}\left(\frac{1}{r}\frac{\partial \psi}{\partial r}\right)\right)\right)+ \delta^2 r\frac{\partial}{\partial r}\left(\frac{1}{r}\frac{\partial}{\partial r}\left(\frac{\partial^2 \psi}{\partial z^2}\right)\right)+ \delta^2 \frac{\partial^2}{\partial z^2}\left(r\frac{\partial}{\partial r}\left(\frac{1}{r}\frac{\partial \psi}{\partial r}\right)\right)+\delta^4 \frac{\partial^4 \psi}{\partial z^4}=0,
\end{equation}
\end{subequations}
where
$$\delta=R/L~~~~~\textrm{and}~~~~~\gamma^2= \phi^{\alpha}_{*}\mu^\alpha/R^2K, $$ 
are dimensionless parameters present in Eqs. (\ref{equ32})-(\ref{equ37}). In this context, $\delta$ refers to the aspect ratio of the atherosclerotic artery. Meanwhile, $\gamma^{2}$ represents the ratio of the viscous drag caused by the growth of the inflammatory tissue phase to the hydraulic resistive force between the constitutive phases of atherosclerotic plaque. \\

\noindent
Now the boundary conditions are rewritten in their dimensionless form as follows (without the dash notation):
\begin{enumerate}[label=(\roman*)]
\item On $r=1$,
\begin{equation}\label{equ38}
v_r^\alpha=0~~~\text{and}~~~v_z^\alpha=0,
\end{equation}
and
\begin{equation}\label{equ38a}
v_r^\beta=0~~~\text{and}~~~v_z^\beta=0.
\end{equation}
\item For some $z=z_0\in[-1,~1]$, and $r=r_0=B(z_{0},t),$
\begin{equation}\label{equ39}
\phi^\alpha v^\alpha_z+\phi^\beta v^\beta_z=v^l_z~~~\text{and}~~~v^l_r=0.
\end{equation}
\noindent
\item The condition (\ref{equ26}) is simplified into both normal and tangential components as follows:
\begin{subequations}\label{equ40}
\begin{equation}\label{equ40a}
-(P-P^l)-\frac{2}{3}\delta^2\gamma^2\left(\frac{1}{r}\frac{\partial(rv_r^\alpha)}{\partial r} + \frac{\partial v_z^\alpha}{\partial z}\right)+ 2\delta^2\left(\gamma^2 \frac{\partial v_r^\alpha}{\partial r}- \beta^2 \frac{\partial v_r^l}{\partial r} \right)=0,
\end{equation}
and
\begin{equation}\label{eqn40b}
\gamma^2 \left(\delta^2 \frac{\partial v_r^\alpha}{\partial z}+ \frac{\partial v_z^\alpha}{\partial r}\right)-\beta^2 \left(\delta^2 \frac{\partial v_r^l}{\partial z}+ \frac{\partial v_z^l}{\partial r}\right)=0,
\end{equation}
respectively. Note that $\beta^{2}$ has the similar interpretation as $\gamma^{2}$ and it can be expressed as $\left(\mu^{l}/\phi^\alpha_{*}\mu^{\alpha}\right)\gamma^{2}$. The expression of $\beta^{2}$ indicates the measure of shear stress impart to the plaque surface caused by the lumenal blood plasma motion. Therefore, $\beta^{2}/\gamma^{2}$ represents the viscosity ratio between the blood plasma and inflammatory tissue phase.
\end{subequations}
\item At $r=a$,
\begin{equation}\label{equ41}
v^l_z=v^p~~~\text{and}~~~v^l_r=0.
\end{equation}
\item The condition (\ref{equ29}) is simplified into
\begin{eqnarray}\label{equ42}
\mathbb{T}^l_{rr}&+&\mathbb{T}^l_{zz}=0, \\\nonumber
\Longrightarrow P^l&+&\beta^2 \delta^2 \left( \frac{\partial v^l_r}{\partial r}+\frac{\partial v^l_z}{\partial z} \right)=0.
\end{eqnarray}
\item Flux Condition: The non-dimensional form of volumetric flow rate is expressed as follows:
\begin{equation}\label{equ43}
\int_0^{a} v^p\,r\,dr+\int_{a}^{B(z,t)} v^l_z\,r\,dr+\int^1_{B(z,t)}\phi^\alpha v^\alpha_z \,r\,dr = 1,
\end{equation}
where $z \in \left[-L,~L\right]$ and $B(z,t)=\frac{d_0}{R}-\frac{\xi(t)}{2}\left(1+\cos(\pi z)\right).$
\end{enumerate}
\subsection{Perturbation Approximation}
\noindent
In this section, we wish to obtain an approximate solution for the boundary value problem formulated by the governing equations and boundary conditions specified in the equations (\ref{equ30})-(\ref{equ41}), using the perturbation methods \cite{pramanik2023two,karmakar2017note,wei2003flow}.
We assume that the aspect ratio ($\delta$) of the region is so small such that $\delta^2 \ll1$ that allows us to apply perturbation approximation. Consequently, with $\delta^2$ as the small parameter, the following variables can be expanded as follows:
\begin{equation}\label{equ44}
(v^i_r,v^i_z,P,\phi^i,\psi)= ((v^i_r)_0,(v^i_z)_0,P_0,\phi^i_0,\psi_0)+ \delta^2 ((u_i)_1,(v_i)_1,P_1,\phi_1^i,\psi_1)+ \textit{O}(\delta ^4),~~\text{for}~~i=\{\alpha,~\beta\}.
\end{equation}
Each variable in (\ref{equ44}) is expressed up to the first order correction, $\delta^2$, as there are no terms of order $\delta$ present in the governing equations and boundary conditions. The non-dimensional governing equations (\ref{equ32})-(\ref{equ37}) and boundary conditions (\ref{equ38})-(\ref{equ43}) are substituted with the expansions (\ref{equ44}). The coefficients of terms corresponding to $\delta^0$ and $\delta^2$ are then equated to zero.
\subsubsection{Determination of Plaque Depth}
\noindent
The temporal behavior of the plaque depth over time can be determined using the plaque boundary equation (\ref{eqn0}). Now, taking the first order partial derivative with respect to $t$:
\begin{equation}\label{equ99}
\frac{\partial r}{\partial t}+\frac{d \xi}{dt} \frac{(1+\cos \pi z)}{2}=0,
\end{equation}
which can be equated with the radial velocity of the inflammatory phase inside the plaque at a point $(r,~z)=(r_0,~z_0)$ on the plaque boundary, to obtain the following first order differential equation
\begin{equation}\label{equ100}
\left.\frac{d\xi}{dt}+\frac{2}{(1+\cos \pi z_0)}v_r^\alpha\right|_{(r,~z)=(r_0,~z_0)}=0.
\end{equation}
Further the expansion (\ref{equ97}) is employed within Eq. (\ref{equ100}):
\begin{equation}\label{equ101}
\frac{d\xi}{dt}+\frac{2}{(1+\cos \pi z_0)}\left[{(v^\alpha_r)}_0\left(r_0=B(z_{0},t)\right)+{\delta^2}{(v^\alpha_r)}_1\left(r_0=B(z_{0},t)\right)+\textit{O}(\delta^4)\right]=0,
\end{equation}
which can be solved numerically to obtain $\xi(t)$ subject to a given initial condition $\xi(0)=\xi_{0}$. However, $\xi(t)$ is obtained as function of $(r_{0},~z_{0})$. Therefore, $\xi(t;z_{0})$ represents the local plaque depth that varies with the location over the plaque.
\subsubsection{Further Assumptions on the Model}
\begin{itemize}
\item The non-inflammatory tissue phase has negligible velocity due to its higher rigidity than the inflammatory tissue phase. The major part of the non-inflammatory tissue phase is the ECM. Some studies on atherosclerosis assume the ECM of plaque to behave as a rigid scaffold \citep{watson2020multiphase}. Therefore, it is logical to consider the non-inflammatory phase as non-motile (i.e., $\mathbf{v}^{\beta}=0$). Mainly, the radial expansion of the plaque depth depends on the radial velocity of the inflammatory tissue phase.
\item We may consider $\gamma^2 \leq 1$ to ensure that the hydraulic resistance force between the inflammatory and non-inflammatory tissue phases dominates the viscous force generated by the growth of the inflammatory tissue phase. $\gamma^{2}$ cannot annihilate the magnitude of $\delta^{2}$ since the above assumption indicates that the product $(\gamma\delta)^{2}\ll1$.
\item We can assume that $\phi^{\alpha}_{0}$ is a constant to simplify the leading order equations for an analytical treatment. At the leading order, we have $\phi^{\alpha}_{0} + \phi^{\beta}_{0} = 1$. Then, for the first and higher order terms of volume fractions, $\phi^{\alpha}_{i} = -\phi^{\beta}_{i}$ when $i\geq1$ is a natural number. However, $\phi^{\alpha}_{i}$'s are spatially and time dependent corresponding to $i\geq1$.
\end{itemize}
The details of the problem at leading order ($\textit{O}(\delta^0)$) and its corresponding solution can be found in Appendix \ref{appendixA} and Appendix \ref{appendixB}, respectively. Similarly, the details of the problem in first order ($\textit{O}(\delta^2)$) and its corresponding solutions are presented in Appendix \ref{appendixC} and Appendix \ref{appendixD}, respectively.
\subsection{Evolution of Cardiac Troponin (cTn)}
\noindent
Cardiac troponin (\textit{cTn}) is a protein released into the bloodstream when there is damage or death to the heart muscle, typically during a heart attack or myocardial infarction. As reported by \citet{mahajan2011interpret}, three clinical cases from the emergency department of a healthcare centre presented with complaints of chest discomfort. The three patients were ages $48$, $54$ and $60$, with a medical history of flu-like symptoms, diabetes mellitus, and heart failure, respectively, alongside the chest discomfort. Cardiac troponin I (cTnI) testing was ordered for the three patients mentioned above. In all three cases, the results were positive, with levels of $0.05$ ng/ml, $0.06$ ng/ml, and $0.06$ ng/ml, respectively, slightly exceeding the diagnostic limit of $0.04$ ng/ml. There is a specific troponin level in the blood. Medical intervention is necessary when it exceeds a certain threshold (used as diagnostic criteria). It can be believed that when heart muscle cells die due to the shortage of oxygen, the neighbouring artery must have undergone a blockage due to plaque formation \citep{clarke2007cell}. An increased blockage in the cardiac artery damages the heart muscle cells and releases more \textit{cTn} into the blood. Through a simple blood test, elevated levels of both the troponins \textrm{cTnI} and \textrm{cTnT} can be measured to determine if the heart has been damaged. Physicians may recommend a \textit{cTn} test if they suspect a patient is experiencing a heart attack or if the patient is already hospitalized\citep{wu2017release,mair2018cardiac}. After a heart attack, \textit{cTn} levels typically normalize within $4$ to $10$ days \citep{wu2017release,mair2018cardiac}. The ultimate goal of us to track the \textit{cTn} elevation for a patient who had a medical history of hyperlipidemia and feeling chest discomfort. As the narrowing of coronary artery leads to inadequate blood supply to the cardiac muscle cells, hence arterial narrowness has a direct relationship with the elevations of \textrm{cTnI} and \textrm{cTnT} \citep{tveit2022cardiac}. Accordingly, this study models the transport of \textit{cTn} within blood and its corresponding generation from the surrounding heart muscle cells. The movement of \textit{cTn} within the bloodstream is believed to be regulated by a non-homogeneous diffusion equation:
\begin{equation}\label{equ102}
D_{\TT}\nabla^2 C^{\TT}+G(t,C^{\TT})=0,
\end{equation}
\noindent
within the blood lumen region of a cardiac artery. Here $C^{\TT}=C^{\TT}(r,z)$ denotes the concentration of both \textit{cTnI} and \textit{cTnT}. This study does not differentiate between them as they may differ in nature, but they do not differ activity-wise. In this regard, \cTn~may be referred to as the local \cTn, which evolved within a stenosed artery. It may primarily be measured at any point within the arterial lumen. $D_{\TT}$ represents the diffusivity of \textit{cTn} and $G(t,C^\TT)$ stands for the net generation of \textit{cTn} as a result of atherosclerotic plaque development with the corresponding expression of $G(t,C^\TT)$ as follows
\begin{equation}\label{equ103}
G(t,C^\TT)= \frac{\lambda_{1}+\left|\left\{\lambda_2 \left({\frac{\xi(t)}{R}}\right)^{n} \mathcal{H}(t-t_1)\right\}/\left\{1+\lambda_3 \left({\frac{\xi(t)}{R}}\right)^{n}\right\} \right| C^\TT}{1+\left|\left\{\lambda_4 \left({\frac{\xi(t)}{R}}\right)^{n} \mathcal{H}(t-t_1)\right\}/\left\{1+\lambda_5 \left({\frac{\xi(t)}{R}}\right)^{n}\right\} \right| C^\TT}-\lambda_6C^\TT\mathcal{H}(t-t_1).
\end{equation}
\noindent
Here, $\lambda_{i}$'s are constants related to the \cTn~generation satisfying the condition $\lambda_{i}\geq0$ for all $i=1,2,..,5$. Among these constants $\lambda_{3}$ and $\lambda_{5}$ are dimensionless. $\lambda_{6}$ represents the decay of {\cTn} while it generates at an increased rate beyond $t=t_{1}$. The concentration of \cTn~within an atherosclerotic artery increases beyond a specified time $t=t_1$, referred to as the reference time for the progression of \cTn~within the bloodstream. Beyond the time level $t=t_{1}$, a hike of \cTn~level within the bloodstream is detected due to the restricted arterial blood flow. $\lambda_{1}$ corresponds to the \cTn~generation within the blood vessel without plaque formation, corresponding to the ambient \cTn~level $0-0.04$ ng/ml. The symbol $\mathcal{H}$ represents the Heaviside function:
\[
\mathcal{H}(t-t_1) =
\begin{cases}
  1,  & \quad t\geq t_{1}\\
  0,  & \quad t < t_{1}
\end{cases}
\]
which can be closely estimated by a smooth function as
\begin{equation}\label{equ104}
\mathcal{H}(x,\epsilon)= \frac{1}{2}\left(1+\frac{2}{\pi}\arctan \frac{x}{\epsilon}\right)~~,~~\epsilon^2 \ll 1.
\end{equation}
\subsubsection{Boundary Conditions for cTn Transport}
\begin{enumerate}[label=(\roman*)]
\item On $r=aR$, we assume that there is no mass flux of \cTn. In other words, no \cTn~diffuses into the pellets as there is no evidence that blood cells may carry \cTn. Hence,
\begin{equation}\label{equ105}
\frac{\partial C^\TT}{\partial r}=0.
\end{equation}
\item At $r=r_0$ (i.e., on the plaque boundary),
\begin{equation}\label{equ106}
C^\TT=C^\TT_{\BB}\left[\lambda_{7}+\left|\frac{\lambda_{8}\left({\frac{\xi(t)}{R}}\right)^{n}}{1+\lambda_{9}\left({\frac{\xi(t)}{R}}\right)^{n}}\right|\right],
\end{equation}
\noindent
where $\lambda_{i+5}$ are constants similar to $\lambda_{1},..,\lambda_{6}$ satisfying the conditions $\lambda_{i+5}\geq0$ for $i=1,2,3$. However, these three are dimensionless. $C^\TT_\BB$ supposed to be the concentration of cTn at the surface of endothelium. The parameter $\lambda_{7}$ regulates cTn such that $\lambda_{7}C^\TT_\BB$ is the concentration of cTn at the endothelium surface in the absence of atherosclerotic plaque growth. Therefore, the entire right hand side of Eq. (\ref{equ106}) indicates cTn concentration at the endothelium surface over the plaque boundary of depth $\xi(t)$. An empirical value of cTn is available in literature with a normal range and it is measured in a unit called nano-gram per millilitre.
\end{enumerate}
\subsubsection{Solution of cTn Transport Equations}
\noindent
We introduce the following transformations $\hat{C}^\TT=C^\TT/C^\TT_\BB,~~\hat{G}=G/\left({{D_{\TT}C^\TT_\BB}/R^2}\right)$ into Eq. (\ref{equ102}), to obtain the corresponding non-dimensional form under the axisymmetry assumption as given by (omitting hat):
\begin{equation}\label{equ107}
\frac{1}{r}\frac{\partial}{\partial r}\left(r \frac{\partial C^\TT}{\partial r} \right)+\delta^2 \frac{\partial^2 C^\TT}{\partial z^2}+ G(t,C^\TT)=0,
\end{equation}
where
\begin{equation}\label{equ108}
G(t,C^\TT)= \frac{\Lambda_{1}+\left|\left\{{\Lambda_2{\xi^{n}(t)}}/{(1+\Lambda_3{\xi^{n}(t)})}\right\}H(t-t_1)\right| C^\TT}{1+\left|\left\{{\Lambda_4\xi^{n}(t)}/{(1+\Lambda_{5}{\xi^{n}(t)})}\right\}H(t-t_1)\right|C^\TT}-\Lambda_6  C^\TT \mathcal{H}(t-t_1).
\end{equation}
\noindent
The boundary conditions corresponding to the reduced transport equation are given by:
\begin{itemize}
\item At $r=a$,
\begin{equation}\label{equ109}
\frac{\partial C^\TT}{\partial r}=0,
\end{equation}
\item At $r=r_0$,
\begin{equation}\label{equ110}
C^\TT= \Lambda_{7}+\left|\frac{\Lambda_{8}{\xi^{n}(t)}}{1+\Lambda_{9}{\xi^{n}(t)}}\right|.
\end{equation}
\end{itemize}
In Eqs. (\ref{equ108}) and (\ref{equ110}), we introduce
\begin{equation}\label{equ111}
\begin{aligned}
&\Lambda_{1}={\lambda_{1}R^2}/{D_{\TT}C^\TT_\BB}, ~~\Lambda_2={\lambda_2R^2}/D_{\TT},~~\Lambda_{3}=\lambda_{3},~~\Lambda_4=\lambda_4C^\TT_\BB,~~\Lambda_{5}=\lambda_{5},\\
&~~\Lambda_6={\lambda_6R^2}/D_{\TT},~~\Lambda_{7}=\lambda_{7},~~\Lambda_{8}=\lambda_{8},~~\Lambda_{9}=\lambda_{9}
\end{aligned}
\end{equation}
\noindent
For simplicity, we set $\Lambda_4$ to zero. As $\delta^2$ is significantly smaller than unity for a cylindrical geometry with small aspect ratio, we neglect the second term in equation (\ref{equ107}). It suggests low axial diffusion rate as compared to the radial diffusion. Consequently, we detect Eq. (\ref{equ107}) as a linear partial differential equation, whose solution can readily be obtained considering the specified boundary conditions (\ref{equ109})-(\ref{equ110}) as given by
\begin{equation}\label{equ112}
C^\TT(r;r_{0},a)=\frac{\Lambda_{1}}{p_1^2}\left[\left(1+\frac{p_1^2}{\Lambda_{1}}(\Lambda_{7}+p_2^2)\right)\left\{\frac{Y_1(p_1a)J_0(p_1r)-J_1(p_1a)Y_0(p_1r)}{Y_1(p_1a)J_0(p_1r_0)-J_1(p_1a)Y_0(p_1r_0)}\right\}-1\right],
\end{equation}
where
$$p_1^2 = \left|\frac{\Lambda_2\xi^{n}(t)}{1+\Lambda_{3}\xi^{n}(t)}-\Lambda_6\right|H(t-t_1)~~~~\textrm{and}~~~~p_2^2 = \left|\frac{\Lambda_{7}\xi^{n}(t)}{1+\Lambda_{8}\xi^{n}(t)}\right|.$$
In addition, $J_{m}$ and $Y_{m}$ stand for Bessel function of first and second kind respectively of order $m\in \mathds{N}\cup\{0\}$. Evidently, $C^{\TT}$ is explicit function of $r$ and implicit function of $t$. Consequently, we may refer to $C^{\TT}$ as the local cTn concentration since it is a function of various lumen points. Therefore, one may integrate $C^{\TT}$ within the plasma region (i.e., $[a,~r_{0}]$) to obtain the average cTn (denoted by $\overline{C}^{\TT}$) which becomes the explicit function of \textit{MPD} $(\xi)$ that varies with $t$. Therefore,
\begin{eqnarray}\label{equ114}\nonumber
\overline{C}^{\TT} &=& 2 \pi \int_a^{r_0}C^\TT r~\dr,\\
&=&\frac{2\pi}{p_1}\left[\mathcal{C}_1\left\{r_0J_1(p_1r_0)-aJ_1(p_1a)\right\}+ \mathcal{C}_2\left\{r_0Y_1(p_1r_0)-aY_1(p_1a) \right\}\right.\\\nonumber
&+& \left.\frac{\Lambda_{1}}{2p_1}\left(a^2-r_0^2\right)\right].
\end{eqnarray}
Both $C^{\TT}$ and $\overline{C}^{\TT}$ play significant role in this study to discuss the probable health complications due to the atherosclerosis. By \cTn~ and \acTn, we will understand local cardiac troponin and average cardiac troponin, respectively.
\section{Results and Discussion}
\subsection{Model Parameterizations}
\noindent
Atherosclerosis is fundamentally a slow process that develop gradually over time. However, there is an increasing acknowledgement that some patients without conventional risk factors may experience an accelerated progression over a few months to $2-3$ years \cite{shah2015rapid}. In this study, we adopt a temporal duration of approximately $2$ years, equivalent to roughly $7\times 10^7$ seconds, to simulate the progression of atherosclerotic plaque. Consequently, the temporal scale of the simulation is expressed into days, with each point on the time scale representing approximately $7$ days. It is also asserted that the principal component of plaque is the inflammatory cell phase, comprising macrophages and foam cells. This postulation allows us to consistently choose the value of $\phi^\alpha_{0}$ and $\phi^\alpha_{*}$ within the range $0.8 \leq \phi^\alpha_{0},~\phi^\alpha_{*} < 1$ throughout our study.\\

\begin{table}[htp]
\centering
\begin{tabular}{ c c c c c }
\hline
Parameter ~~~~~~& Dimensionless value or Range ~~~~& Supporting literature \\
\hline
$d_0$ & $0.865$ & \citet{wang2019hydraulic} \\
$\gamma^2$ & $1.8\times 10^{-3}$ & \citet{wang2019hydraulic} \\
$\delta$ & $0.03 \leq \delta \leq 0.3$ & \citet{wang2019hydraulic,al2000coronary} \\
$a$ & $1.0\times 10^{-2}\leq a \leq 1.0\times 10^{-1}$ & Considered \\
$\phi^\alpha_{0}$ & $0.9$ & Considered \\
$\phi^\alpha_{*}$ & $0.9$ & Considered \\
\hline
\end{tabular}
\caption{Calculated non-dimensional parameter values extracted from the above-mentioned literature.}
\label{table:1}
\end{table}
\noindent
Arteries come in different sizes, ranging from the smallest arterioles with a diameter of $0.01-0.30$ mm to the largest arteries, such as the aorta, with a diameter of $10-25$ mm \cite{wahood2022radial}. The hydraulic permeability of the artery wall is $1.3438 \times 10^{-18} \text{m}^2$, as reported by \citet{wang2019hydraulic,wang2014enhanced}. To calculate the hydraulic resistivity of the arterial wall, it is crucial to know the viscosity of atherosclerotic plaque built up within the intima layer of the artery wall. There is a requirement for further literature to establish precise values of $\mu^\alpha$, representing the viscosity of the inflammatory cell phase within the plaque. Arterial plaque primarily consists of fat and lipids. Therefore, it is reasonable to assume that the value of the parameter $\mu^\alpha$ is equivalent to the viscosity of lipids. \citet{adrien2022best} provides the viscosity value of the bilayer lipid, which is $243 \times 10^{-3} \text{kg m}^{-1} \text{s}^{-1}$. On the other hand, the viscosity of blood plasma is found in literature $1.2-1.3\times10^{-3}$ $\textrm{kg}\textrm{m}^{-1}\textrm{s}^{-1}$ \cite{nader2019blood} and this range of values is much lesser than that of lipid as reported by \citet{adrien2022best}. Therefore, the order of magnitude of viscosity of the inflammatory cell phase within the plaque can significantly exceed that of the blood within the lumen. Without loss of generality, we are allowed to consider $\mu^l/\mu^\alpha < 1$.\\

\noindent
Now, we are in a position to compute a magnitude range for $\gamma^{2}$ which is proportional to Darcy's constant (or Darcy's number), rendering the permeability inside plaque material. In this regard, the magnitude of $\gamma^{2}$ is considered within the range $\left[10^{-4},~10^{-3}\right]$ since in standard literature, the magnitude of Darcy's constant is obtained within the said interval. However, to find a realistic value of Darcy's constant, appropriate literature should be examined to forecast that vascular wall permeability is predominant. On the other hand, $\beta^{2}$ has a similar interpretation as $\gamma^{2}$. We notice that $\mu^{l}/\mu^{\alpha}<1$ is equivalent to $\beta^2/\gamma^2 < 1$. Therefore, $\beta^{2}$ is expected to belong to the interval $\left[10^{-7},~10^{-4}\right]$. Following this criterion, we have chosen four specific values for $\beta^2/\gamma^2$, which are $10^{-1}, 10^{-2}, 10^{-3},$ and $10^{-4}$. The parameter $a$ regulates \textit{PRR}. White blood cells (WBCs) are typically $10-20$ $\mu$m in diameter. They are about three times larger than red blood cells (RBCs), which are roughly $6-8$ $\mu$m in size \cite{schmid1980morphometry}. In this study, it is important to choose the highest possible value of $a$ less than $|d_{0}-\xi(t)|$ to avoid any obstruction of blood flow. According to \citet{wang2019hydraulic}, $d_{0}=0.865$. Purposefully, we choose $a\in[0.01,~0.1]$ in order to include a large varieties of the pellets. Lastly, the length of the atherosclerotic plaque, denoted as $L$, can be estimated by considering it roughly equivalent to the size of an artery stent, a medical device employed in treating narrowed or obstructed arteries. This approximation situates the plaque's length within the range of $8 \text{mm}$ to $38 \text{mm}$, as reported by \citet{al2000coronary}. Based on the range for $L$ and $R$ mentioned above, the perturbation quantity $\delta$ is computed to belong within the range $0.03\leq\delta\leq0.3$. Based on the above discussions, we mention the respective non-dimensional values for the parameters $d_0$, $a$, $\phi^\alpha_{0}$, $\phi^\alpha_{*}$, $\gamma$, and $\delta$, as presented in Table \ref{table:1}. \\

\begin{table}[htp]
\centering
\begin{tabular}{ c c }
\hline
Parameter ~~~&~~~ Value \\
\hline
$\Lambda_1$ & ~~~~$1\times10^{-2}$ \\
$\Lambda_2$ & ~~~~$1$ \\
$\Lambda_3$ & ~~~~$1\times10^{-2}$ \\
$\Lambda_6$ & ~~~~$1\times10^{-2}$ \\
$\Lambda_7$ & ~~~~$1\times10^{-3}$ \\
$\Lambda_8$ & ~~~~$1$ \\
$\Lambda_9$ & ~~~~$1\times10^{-1}$ \\
\hline
\end{tabular}
\caption{Estimated values of the dimensionless parameters involved in cardiac troponin (cTn) generation term.}
\label{table:2}
\end{table}
\noindent
In this paragraph, we analyze the evolution of cardiac troponin (\cTn) in terms of parameters $\Lambda_{i}$ for $i=1,2,3,6,7,8,9$ (presented in Table \ref{table:2}). Initially, we set the value of the parameter $\Lambda_1$ as equal to $1\times10^{-2}$. At the beginning of atherosclerosis, it is evident that the plaque depth becomes zero. Utilizing this observation in the boundary condition (\ref{equ110}) corresponding to the \cTn~ concentration gives $C^\TT= \Lambda_{7}$. Furthermore, considering the normal range of \cTn~concentration in the blood, which lies between $0$ and $0.04$ ng/ml, we make a reasonable choice by setting $\Lambda_{7} = 1\times10^{-3}$. The plaque depth increases with time  leading to an elevation in \cTn~ concentration in the blood. As a result, the parameters $\Lambda_2$ in Eq. (\ref{equ108}) and $\Lambda_7$ in (\ref{equ110}) are assumed to be greater than zero. In particular, we set $\Lambda_2=\Lambda_8=1$. For the sake of brevity, the parameters $\Lambda_3$, $\Lambda_6$ and $\Lambda_8$ are set to $0.01$, $0.01$ and $0.1$, respectively. Here, the variable $t_1$ denotes the specific time when the \textit{cTn} concentration starts to increase, indicating that prior to $t=t_1$, the \cTn~ concentration remains within the normal range, and after $t=t_1$ the concentration begins to rise due to plaque growth. We manually set a time $t=t_{1}$ in the present investigation such that beyond which a gradual growth in local \cTn~is noticed with increased plaque depth. We set $t_{1}=40$ which is equivalent to $280$ days and this value of $t_{1}$ is used further to generate graphical results in our study.
\subsection{Various Aspects of Plaque Growth}
\begin{figure}[htp]
\centering
\begin{subfigure}[b]{0.49\textwidth}
\centering
\includegraphics[width=\textwidth]{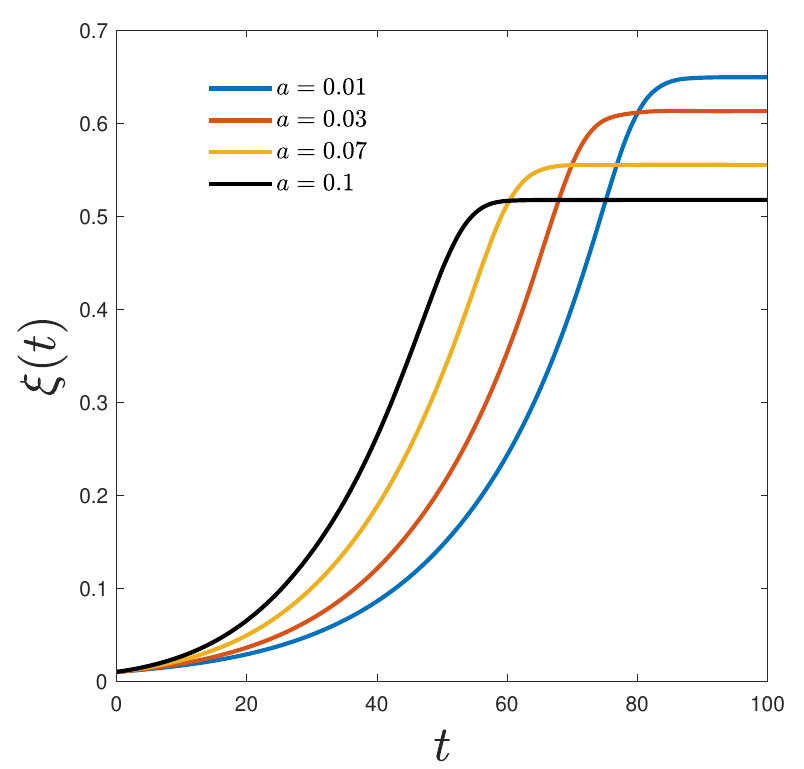}
\caption{}
\label{fig:4a}
\end{subfigure}
\begin{subfigure}[b]{0.49\textwidth}
\centering
\includegraphics[width=\textwidth]{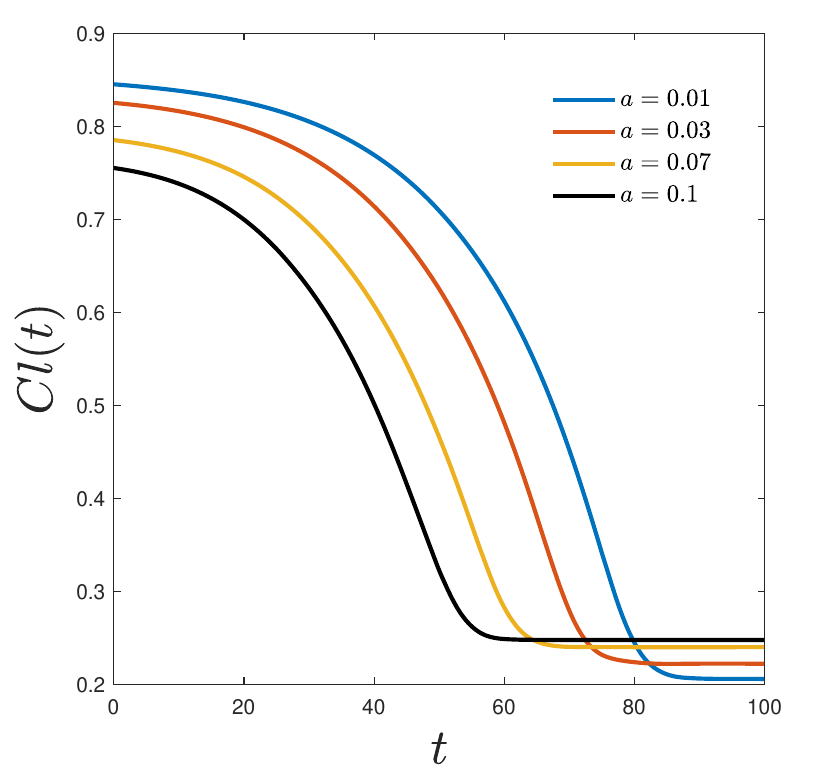}
\caption{}
\label{fig:4b}
\end{subfigure}
\caption{(a) Evolution of  plaque (stenosis) depth and (b) reduction of clearance depth over time corresponding to $a=1 \times10^{-3},~2.58\times10^{-2},~ 5.05\times10^{-2},~7.53\times10^{-2}$ and $1 \times10^{-1}$}
\end{figure}

\noindent
The temporal dynamics of plaque growth are illustrated in Figure \ref{fig:4a} in terms of $\xi(t)$ (\textit{MPD} varies with time) corresponding to different \textit{PRR} values, considering an axial location $z=z_0$ along the plaque boundary. The temporal progression of plaque growth initially (or in other words, \textit{MPD}) follows an exponential growth that diminishes over time, resulting in a stable state. Therefore, the rate at which the plaque boundary progresses remains nearly constant after a threshold time. In advanced stages, the plaque growth rate becomes zero due to the normal stress exerted by the flowing blood plasma and cells within the lumen at various locations of the plaque surface. The plaque growth progression over time is examined for four different values of $a$ \emph{viz} $1\times 10^{-2},~3\times 10^{-2},~7\times 10^{-2},~1\times 10^{-1}$. For the sake of convenience, we level above four different \textit{PRR} as $a_{1}$, $a_{2}$, $a_{3}$ and $a_{4}$ in respective manner. These observations suggest that the growth graph exhibits a progressively rising behavior with $a$.\\

\noindent
For two different values of $a$: \emph{viz.} $a=a_{4}$ and $a=a_{2}$, with $a_{4}(=0.1)>a_{2}(=0.03)$ a higher growth rate is observed for $a=a_{4}$ as compared to $a=a_{2}$, quickly reaches to the steady state for $\xi(t)$ when $a=a_{4}$. The maximum value (or static value) of $\xi(t)$ is larger for $a=a_{2}$ despite its slower growth compared to $a=a_{4}$. The maximum static magnitudes of $\xi(t)$ for $a_{4}$ and $a_{2}$ are denoted as $\xi_{\textrm{max},a_{4}}$ and $\xi_{\textrm{max},a_{2}}$ respectively, occurring at times $t_{a_{4}}$ and $t_{a_{2}}$. We note from Figure \ref{fig:4a} that $\xi_{\textrm{max},a_{2}}>\xi_{\textrm{max},a_{4}}$. The growing plaque experiences the low radial stress from the lumen when $a=a_{2}$ compared to $a=a_{4}$ (\textit{PRR} value becomes more extensive). On the other hand, the more considerable radial expansion of the plaque becomes possible when the plasma region becomes wider. However, during the expansion of plaque, a stage is reached (in these cases, $t=t_{a_{i}}$ for $i=1,2,3,4$) when the generated radial stress from the lumen can halt the plaque expansion as on the plaque-lumen interface, the normal stress balance condition is assumed. Consequently, $\xi(t)$ takes a static value rapidly when $a=a_{4}$ and a thinner plug region allow plaque growth to continue for an extended period. Note that one cannot accurately depict the \textit{PRR} values due to the arbitrary nature of the pellet sizes and their count.\\

\noindent
The variation of \textit{MPD} shows a significant dependency on the radius $a$, which enables us to think about the plasma region, which provides a space for plaque expansion. Hence, it is essential to identify a relevant function. Consequently, through Figure \ref{fig:4b}, we analyze the temporal changes in Lumen Clearance ($CL$), defined in the dimensionless form as follows
\begin{equation}\label{equ120}
Cl(t)=\frac{1}{R}(d_{0}-aR-\xi(t)),
\end{equation}
\noindent
which is the minimum available space for the motion of blood plasma within the artery lumen when plaque development occurs. This lumen clearance ($CL$) region enables the smooth passage of pellets and provides adequate space for plaque development. Hence, we must simultaneously discuss the beneficial and detrimental aspects of the lumen clearance (\textit{CL}). There, a temporal reduction of \textit{CL} is illustrated, demonstrating an initial phase of exponential decay followed by a steady value for the subsequent period. The \textit{CL} converges to a stable state, indicating that the rate of decrease remains nearly constant after the initial stage. This phenomenon contrasts with the relationship observed between \textit{MPD} and time. However, the stability may be compromised in the presence of another advanced plaque. In advanced stages, some cytokines such as interferon-$\gamma$ (INF$\gamma$) and matrix metalloproteinases (MMP) may cause the fibrous cap to rupture, exposing thrombogenic plaque material to the bloodstream. This incident can lead to a complete occlusion of the \textit{CL} space. In contrast to the behavior displayed by \textit{MPD} when $a=a_{i}~(i=1,2,3,4)$ with $a_{i}>a_{j}$ and $i>j$, we observe that the \textit{CL} experiences more pronounced temporal depletion with the same values of $a$ taken in the same order of magnitude. While there is a considerable difference in the peaks of $MPD$ for the four values of $a$, the variations in the troughs of \textit{CL} are less significant for the same four values of $a$. As the \textit{CL} is the minimum available space for the blood plasma to flow within an atherosclerotic artery, the impact of the \textit{PRR} values becomes marginal. We earlier stated that blood cells (pellets) maintain a specific size. Hence, we have to discuss the arterial clearance, excluding the plug region. The overall analysis utilize the parameters specified in Table \ref{table:1}.\\

\noindent
Understanding how blood viscosity and plaque material viscosity affect \textit{MPD} and \textit{CL} is crucial. Physicians often recommend blood thinners for patients with atherosclerosis. The impact of blood viscosity on \textit{MPD} and \textit{CL} is demonstrated in Table \ref{table:4}. When keeping $\gamma^2$ fixed, a decrease in the ratio of $\beta^2/\gamma^2$ indicates a reduction in blood viscosity, leading to a decrease in shear stress. Consequently, we observe an increase in plaque growth and a decrease in clearance (third and fourth columns of Table \ref{table:4}, respectively). The fifth column illustrates the three differences in the values of $\beta^{2}/\gamma^{2}$, and the sixth column shows the corresponding changes in \textit{MPD} and \textit{CL}. The last column displays their percentage change. With each difference in the value of $\beta^{2}/\gamma^{2}$, there is a rapid increase in $MPD$ and a decrease in $CL$ by the same amount. This tabular demonstration indicates reduced shear stress, leading to increased plaque growth and decreased clearance. Therefore, it is crucial to monitor the change in shear stress at the plaque boundary caused by the movement of plasma as the plaque grows. \\

\begin{table}[ht!]
\centering
\begin{tabular}{ p{1.5cm} p{1.5cm} p{1.5cm} p{1.5cm} p{3.5cm} p{2.5cm} p{2.3cm} }
\hline
Sl No. & $\beta^2/\gamma^2$ & Mean $\xi(t)$ & Mean $Cl(t)$ & Difference in $\beta^2/\gamma^2$ & Change in Mean $\xi(t)$ and $Cl(t)$ &~\% change in $\xi(t)$ and $Cl(t)$\\
\hline
$1$ & $10^{-1}$ & 0.1849 &  0.6791 &                               &          &       \\
    &           &        &         &$\left|10^{-1}-10^{-2}\right|$ & ~~0.0693 & ~~ 77  \\
$2$ & $10^{-2}$ & 0.2542 &  0.6098 &                               &          &       \\
    &           &        &         &$\left|10^{-2}-10^{-3}\right|$ & ~~0.0223 & ~~248 \\
$3$ & $10^{-3}$ & 0.2765 &  0.5875 &                               &          &       \\
    &           &        &         &$\left|10^{-3}-10^{-4}\right|$ & ~~0.0081 & ~~900 \\
$4$ & $10^{-4}$ & 0.2846 &  0.5794 &                               &          &       \\
\hline
\end{tabular}
\caption{Percentage increase and decrease in $\xi(t)$ and $Cl(t)$ respectively with respect to $\beta^2/\gamma^2$.}
\label{table:4}
\end{table}
\noindent
Wall shear stress refers to the tangential force exerted by the flowing fluid, typically blood, on the surface of the arterial wall \cite{sun2007effects}. It is well-established that vessel segments characterized by low wall shear stress or highly oscillatory wall shear stress are at the highest risk for atherosclerotic plaque development \cite{sun2007effects}. It is observed from the research by \citet{shaaban2000wall} that low wall shear stress plays a crucial role in the accumulation of LDL or elevation of LDL concentration (Low-Density Lipoprotein) within the arterial wall by facilitating the passage of LDL through the various wall layers. This accumulation leads to the formation of arterial plaque.\\

\begin{figure}[htp]
\centering
\begin{subfigure}[b]{0.49\textwidth}
\centering
\includegraphics[width=\textwidth]{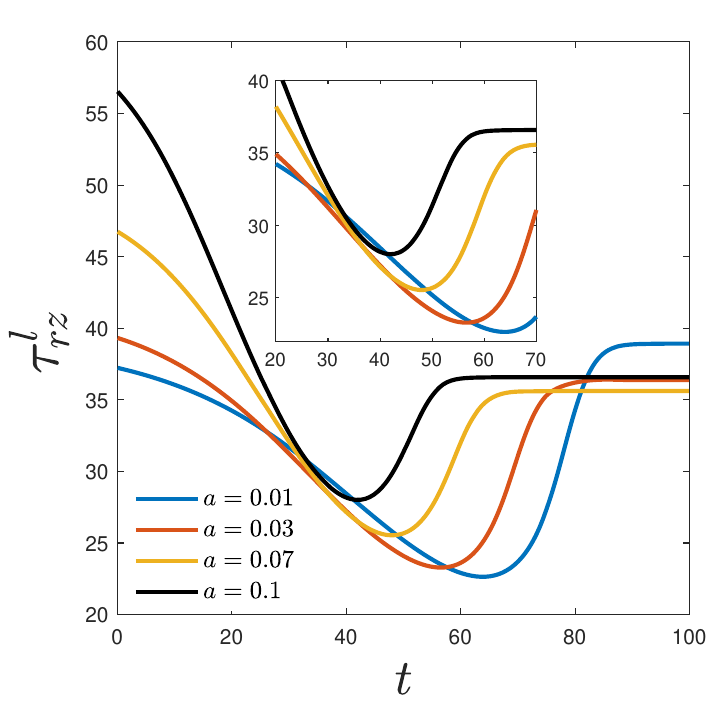}
\caption{}
\label{fig:9a}
\end{subfigure}
\begin{subfigure}[b]{0.49\textwidth}
\centering
\includegraphics[width=\textwidth]{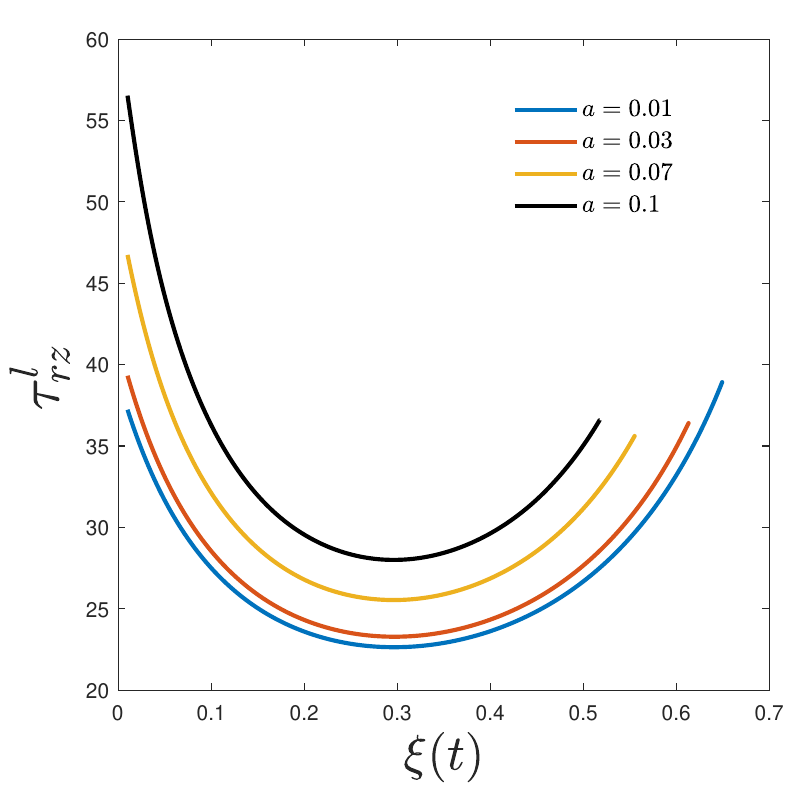}
\caption{}
\label{fig:9b}
\end{subfigure}
\caption{Variation of stress (a) with time for $a=1 \times10^{-2},~3\times10^{-2},~ 7\times10^{-2}$,~and $1 \times10^{-1}$, (b) with plaque length for various $a=1 \times10^{-2},~3\times10^{-2},~ 7\times10^{-2}$,~and $1 \times10^{-1}$}
\end{figure}

\noindent
Figure \ref{fig:9a} depicts the change in the temporal behavior of the shear stress $\left(\tau^{l}_{rz}\right)$ exerted by the blood plasma on the plaque boundary at a specific location $r=r_0$, considering different \textit{PRR} values. The profile of $\tau^{l}_{rz}$ initially exhibits a decreasing behavior, followed by a subsequent rising pattern, and ultimately reaches a stable state. On the other hand, Figure \ref{fig:9b} illustrates the variation of $\tau^{l}_{rz}$ at the plaque boundary with the maximum depth of atherosclerotic plaque at any instant. Each curve shows a concave shape that corresponds to \textit{PRR}. Specifically, as the plaque depth increases, the $\tau^{l}_{rz}$ initially decreases to a particular value and then increases. The larger \textit{PRR} corresponds to higher shear stress in both instances. Furthermore, it is evident that low shear stress is a significant factor contributing to the development of plaque growth. At first, the speed of blood plasma is slowed down by the resistance caused by the growing plaque. Moreover, such a chance of blocking becomes even more pronounced with the increasing \textit{PRR} values. Then, the plasma flow speeds up as it tries to pass through the narrowest point (where \textit{MPD} is at its maximum). The velocity gradient of blood plasma in the axial direction reduces as the resistance from the growing plaque enhances, highlighting the impact of plaque growth on the plasma motion. As soon as it overcomes the resistance, a further increment in the velocity gradient is resulted. This phenomenon is familiar in the context of Newtonian fluid mechanics where axial velocity gradient is proportional to the shear stress. In this study, the blood plasma is assumed to show the Newtonian behavior. Besides, this study assumes that the pellets move at the same speed as the plasma, following the boundary condition (\ref{equ27}). Consequently, the contribution of the movement of the pellets towards the stress field likely to display a similar behavior with plasma. We further analyze the effect of the pellet velocity with time. \\

\noindent
Figure \ref{fig:9a} enables us to locate a particular instant $t=35$ when all the descending profiles of $\tau^{l}_{rz}$ coincide. The $\tau^{l}_{rz}$ curve corresponding to $a=0.1$ changes direction, becoming ascending and finally reaching a stable value over time. Therefore, $\tau^{l}_{rz}$ approaches a minimum around $t=35$ when $a=0.1$. Other profiles of $\tau^{l}_{rz}$ correspond to $a=0.07$, $a=0.03$, and $a=0.01$, showing similar behavior but with delayed times to reach the minimum. Corresponding to the plug region with a smaller diameter (smaller \textit{PRR} values), a higher chance is to obtain the minimum $\tau^{l}_{rz}$. In the case of $a=0.01$, the most delay is observed to attend the minimum value of $\tau^{l}_{rz}$ compared to the other three cases. The increased \textit{PRR} values corresponds to a larger volume of blood cells, leading to elevated shear stress and momentum transfer to the plaque surface through the lumen. The rate of plaque growth increases with reduced shear stress, promoting plaque development. In this scenario, blood-borne low-density lipoproteins (LDLs) and other trans-lipids might penetrate the plaque, leading to the continuation of the growth process. It is widely recognized that low shear stress plays a role in plaque development, supporting our research findings. On the other hand, increased shear stress might slow down plaque growth but can cause rapture of the plaque surface, which attracts red blood cells to accumulate at the damaged portion. This incident can lead to a complete block of an artery.\\

\begin{figure}[htp]
\centering
\includegraphics[width=0.49\textwidth]{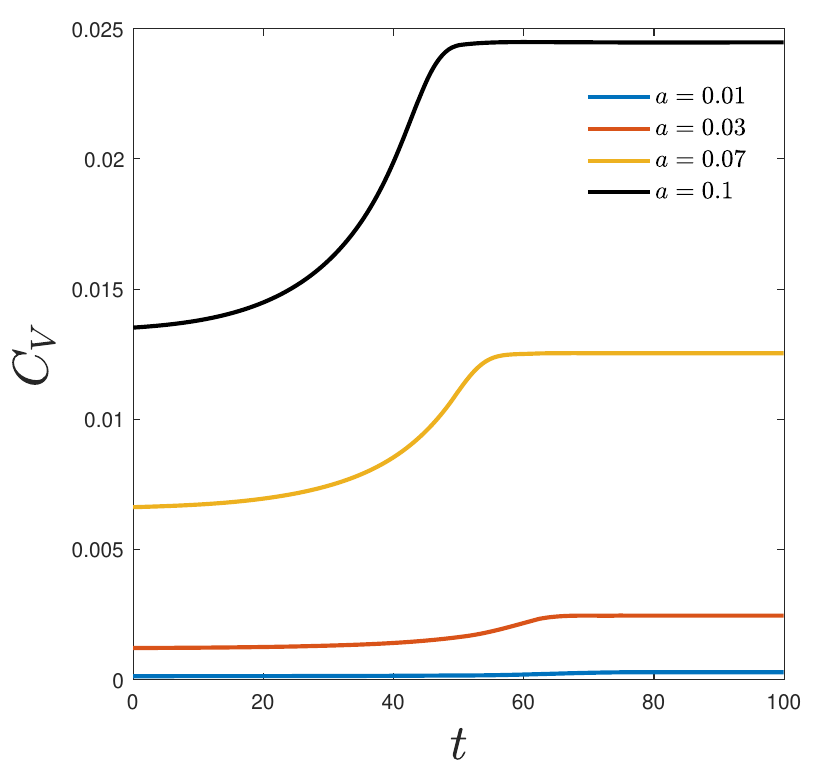}
\caption{Variation of \textit{CVF} $\left(C_{\text{v}}\right)$ with time for $a=1 \times10^{-2},~3\times10^{-2},~ 7\times10^{-2}$,~and $1 \times10^{-1}$.}
\label{fig:10a}
\end{figure}

\begin{figure}[htp]
\centering
\begin{subfigure}[b]{0.49\textwidth}
\centering
\includegraphics[width=\textwidth]{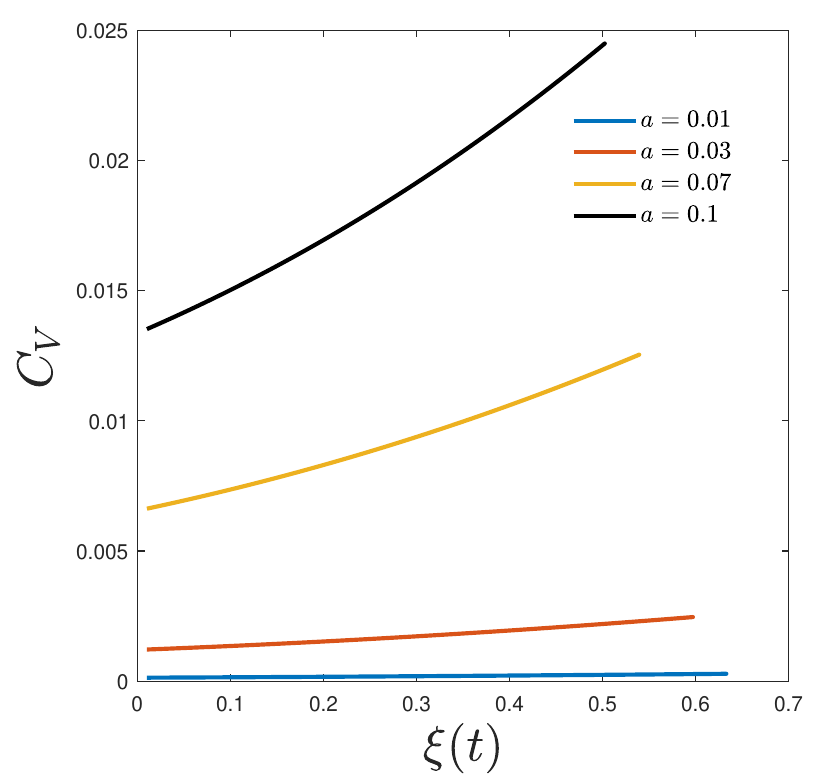}
\caption{}
\label{fig:11a}
\end{subfigure}
\begin{subfigure}[b]{0.49\textwidth}
\centering
\includegraphics[width=\textwidth]{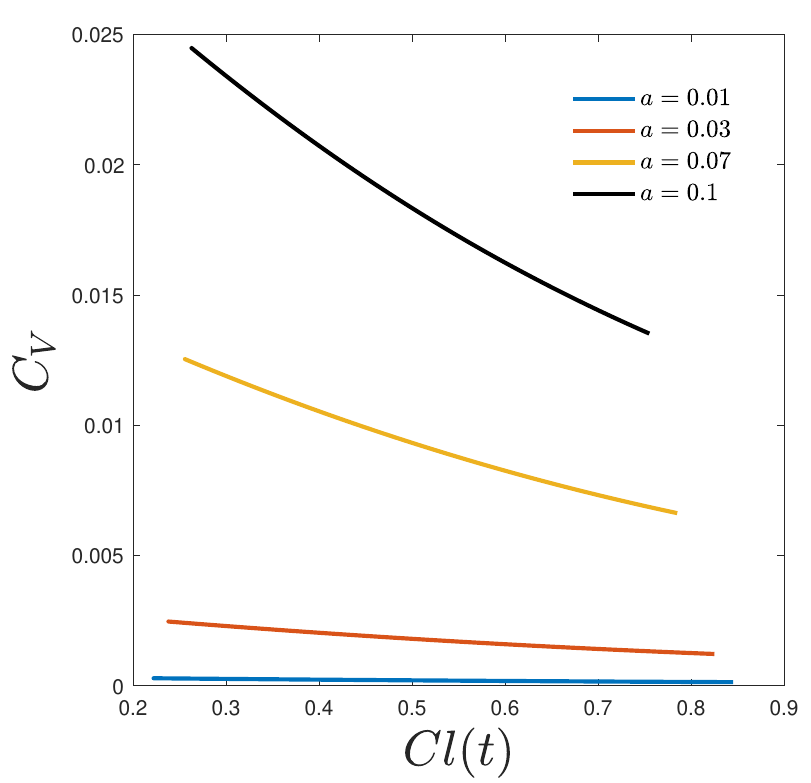}
\caption{}
\label{fig:11b}
\end{subfigure}
\caption{Variation of CVF $\left(C_{\text{v}}\right)$ with (a) \textit{MPD} (b) clearance for $a=1 \times10^{-2},~3\times10^{-2},~ 7\times10^{-2}$,~and $1 \times10^{-1}$.}
\end{figure}

\noindent
It has been established that \textit{MPD} and \textit{CL} are local variables that depend on $z_{0}$. Therefore, we need something that does not depend on any specific point on the plaque boundary. In this context, we must introduce something similar to Hematocrit, which is defined as the percentage by volume of red cells in the blood. In this investigation, we may define a quantity called the \emph{Cell Volume Fraction} (\textit{CVF}) in blood:
\begin{eqnarray*}
C_{v}&=& \frac{\text{Volume of Plug region}}{\text{Volume of the Entire Lumen}} \\
&=&  \frac{2 \pi a^2}{2 \pi \int_{-1}^{1} \int_{0}^{B(z,t)} r \,dr \,dz} = \frac{2 a^2}{\frac{3}{4}\left(\xi(t)\right)^2-\frac{2 d_0}{R}\xi(t)+2\left(\frac{d_0}{R}\right)^2}.
\end{eqnarray*}
\noindent
The Figures \ref{fig:10a}, and \ref{fig:11a}-\ref{fig:11b} describe the variations in \textit{CVF} influenced by $t$, \textit{MPD} and \textit{CL} respectively, with consideration of various values of \textit{PRR} ($a$) into account. The evolution of \textit{CVF} shows an initial phase of monotonic growth followed by a subsequent phase wherein the growth rate diminishes to a stable equilibrium state. This incident happens because, as time passes, the buildup of plaque reduces blood flow, causing blood cells to accumulate in the plaque area and the \textit{CVF} to increase. Later, \textit{CVF} stabilizes once \textit{MPD} reaches its maximum. Moreover, it shows a similar pattern of variation with $t$ as observed in case of Haematocrit. As earlier, four different values of $a$ namely $1 \times10^{-2},~3 \times10^{-2},~7 \times10^{-2},~1\times10^{-1}$ are considered here. When the value of \textit{PRR} is smaller (when the plug region accommodates a smaller amount of blood cells), the temporal variation in \textit{CVF} is found to be insignificant; however, for larger values, this variation becomes significant.\\

\noindent
In other words, an expansion in the volume of the plug region indicates a concurrent rise in \textit{CVF}. Next, we see a significant increase in \textit{CVF} across \textit{MPD} in Figure \ref{fig:11a}. However, in Figure \ref{fig:11b}, we observe a contrasting behavior as \textit{CVF} experiences a notable decrease with \textit{CL}. As depicted in Figure \ref{fig:11a}, \textit{CVF} exhibits an enhancement correlating with the increasing plaque depth following the previous discussion. Further note that with the increase in $a$, the plug region expands in volume, signalling a concurrent elevation in \textit{CVF}. In addition, the highest value of \textit{CVF} reaching $0.025$, is obtained close to $\xi(t)=0.5$ for the largest size of \textit{PRR} considered here, i.e., $a=0.1$. In contrast, Figure \ref{fig:11b} shows \textit{CVF} decreases with increased clearance, presenting a behavior opposite to Figure \ref{fig:11a}.
\subsection{Temporal variation of Cardiac Troponin (\textit{cTn})}
\begin{figure}[htp]
\centering
\begin{subfigure}[b]{0.485\textwidth}
\centering
\includegraphics[width=\textwidth]{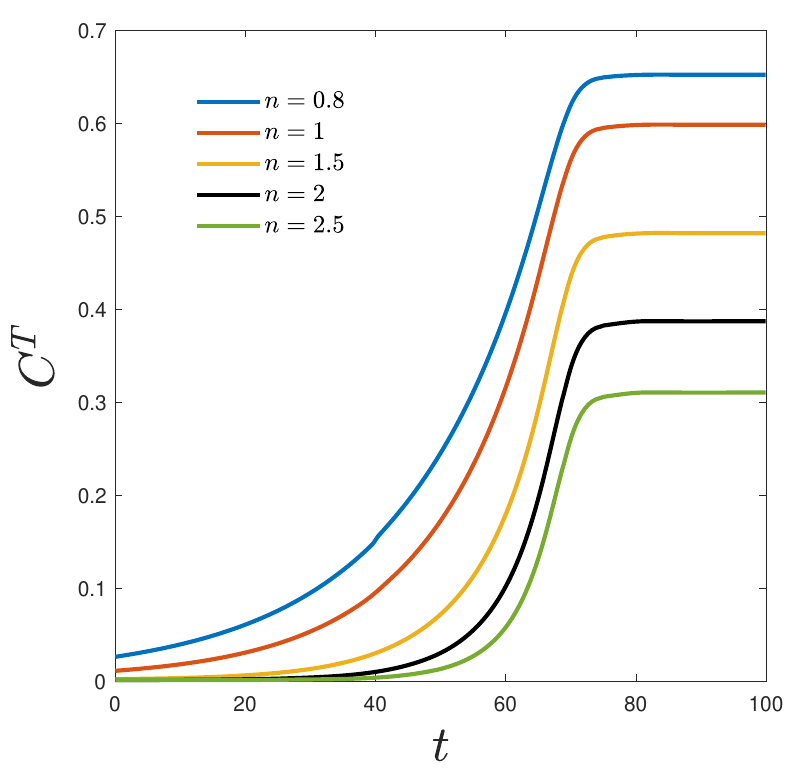}
\caption{}
\label{fig:5a}
\end{subfigure}
\begin{subfigure}[b]{0.5\textwidth}
\centering
\includegraphics[width=\textwidth]{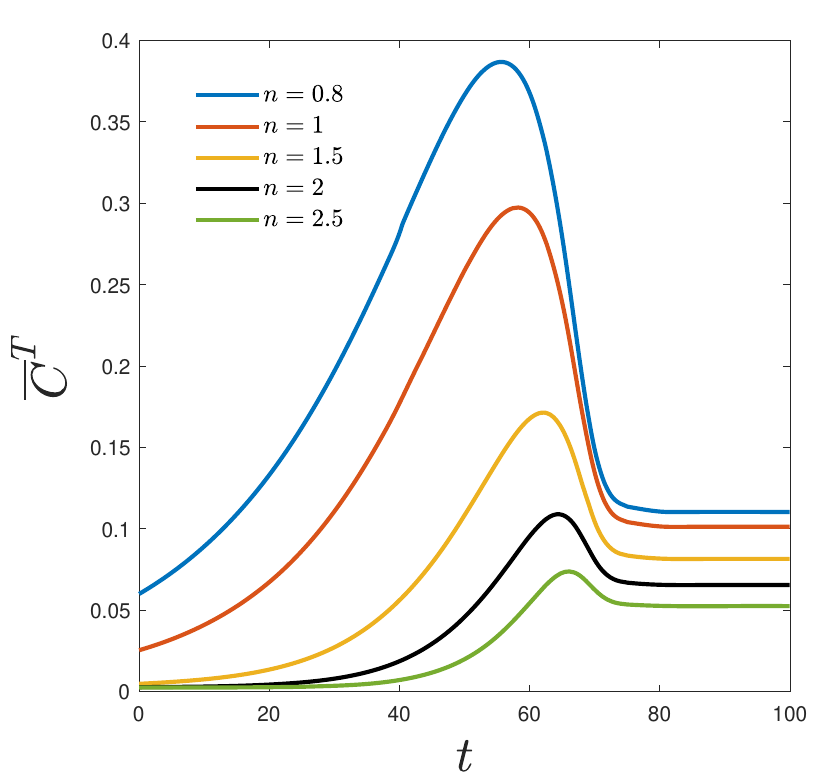}
\caption{}
\label{fig:5b}
\end{subfigure}
\begin{subfigure}[b]{0.5\textwidth}
\centering
\includegraphics[width=\textwidth]{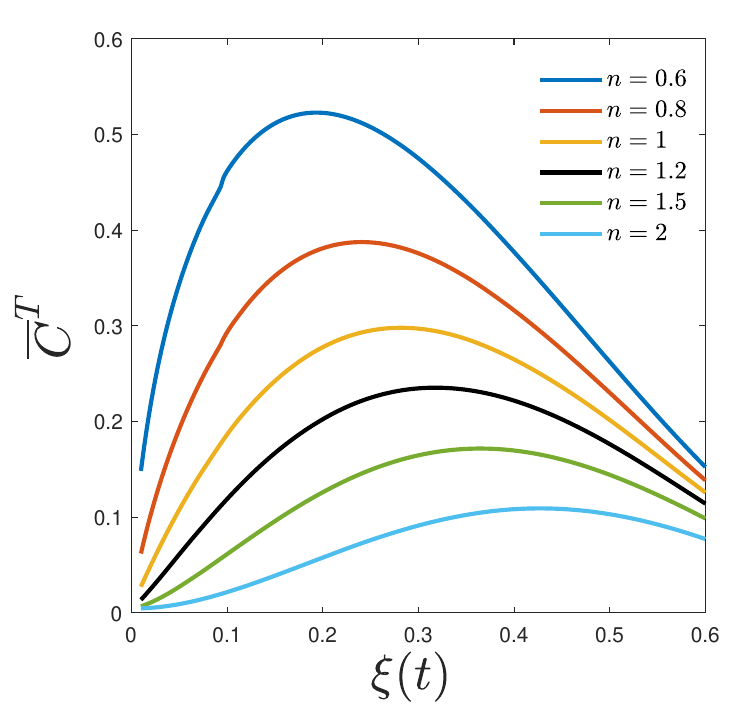}
\caption{}
\label{fig:5c}
\end{subfigure}
\caption{Evolution of (a) cardiac troponin (cTn) concentration with time (b) average cardiac troponin ($\overline{\textit{cTn}}$) concentration with time and (c) average cardiac troponin concentration ($\overline{\textit{cTn}}$) with $\xi(t)$, corresponding to $n = 0.5,~0.8,~1,~1.2,~1.5,~2$.}
\end{figure}

\noindent
As plaque growth progresses, it gradually lowers the clearance of the lumen and initiates obstruction to the motion of various blood cells within the artery. Red blood cells (RBCs) deliver oxygen to the tissues. Consequently, obstruction to the motion of RBCs leads to a decrease in the blood oxygen level that further causes potential damage to the heart muscle—the damaged heart muscle release cardiac troponin (\textit{cTn}) into the bloodstream. The normal range in the blood falls from $0$ to $0.04$ ng/mL. To accommodate this, we consider the generation term for \textit{cTn} in equation (\ref{equ102}) such that the increase in \textit{cTn} concentration begins after a defined time $t_1$, which is assumed from some experimental studies \cite{mahajan2011interpret,park2017cardiac}. In Figures \ref{fig:5a}-\ref{fig:5b}, we analyze the temporal variations in local troponin concentration (\textit{cTn}) at a particular point in the \textit{CL} area and average \cTn~concentration $\overline{\textit{cTn}}$, respectively, for five different values of $n$.
\noindent
The \textit{cTn} attributes to the state of \textit{cTn} generation inside an atherosclerotic artery at some arbitrary point within the \textit{CL} region/plasma region. The maximum value of \textit{cTn} concentration may be location dependent. Therefore, we must focus on the average troponin concentration ($\overline{\textit{cTn}}$) which forecasts the maximum elevation of \textit{cTn} over the whole atherosclerotic artery. However, the $\overline{\textit{cTn}}$ depends on a point $z_{0}$ located over the plaque boundary. Hence, we try to prognosticate the issue of geometry dependency of the \textit{cTn} distribution. The exponent $n$ corresponds to the rapidity in the troponin generation. It is evident that both the troponin concentrations initially remain near zero. Starting with an increasing trend over time, \textit{cTn} eventually stabilizes at due course of time. On the other hand, the $\overline{\textit{cTn}}$ increases over time to reach a maximum and then decreases to a constant value. Furthermore, as time progresses, both concentrations diminish with the rising exponent $n$. The exponent $n$ decides how fast the \textit{cTn} concentration increases over time within an artery undergoing atherosclerosis. Without a suitable mathematical correlation between \textit{MPD} and \textit{cTn} concentration, this study proposes a generation function of \textit{cTn} dependent on plaque depth. The pronounced effect of plaque depth over the $\overline{\textit{cTn}}$ can be observed as $n\leq1$, whence such effect becomes less remarkable for $n>1$. An immense value of the exponent $n$ subdues the impact of \textit{MPD} as suggested through the structure of the \textit{cTn} source term in Eq. (\ref{equ107}). Figure (\ref{fig:5c}) supports the above observation. The arterial cross-section depends on time $t$, and the cosinusoidal geometry of the plaque stands responsible for the nature of $\overline{\textit{cTn}}$ as shown in Figures (\ref{fig:5b}) and (\ref{fig:5c}) concerning $t$ and $\textit{MPD}$ respectively.\\

\begin{figure}[htp]
\centering
\begin{subfigure}[b]{0.49\textwidth}
\centering
\includegraphics[width=\textwidth]{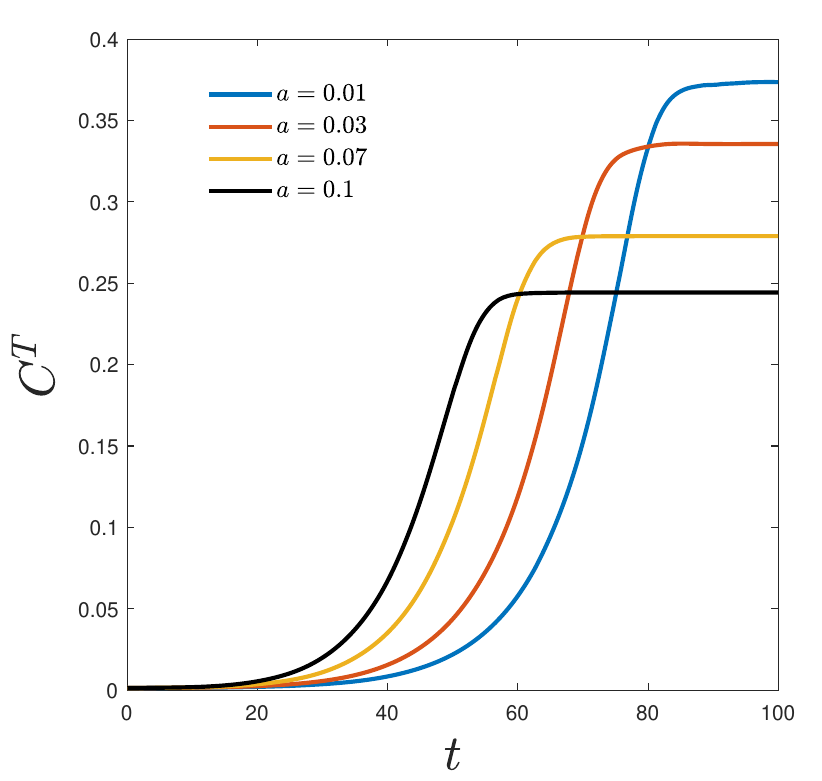}
\caption{}
\label{fig:6a}
\end{subfigure}
\begin{subfigure}[b]{0.5\textwidth}
\centering
\includegraphics[width=\textwidth]{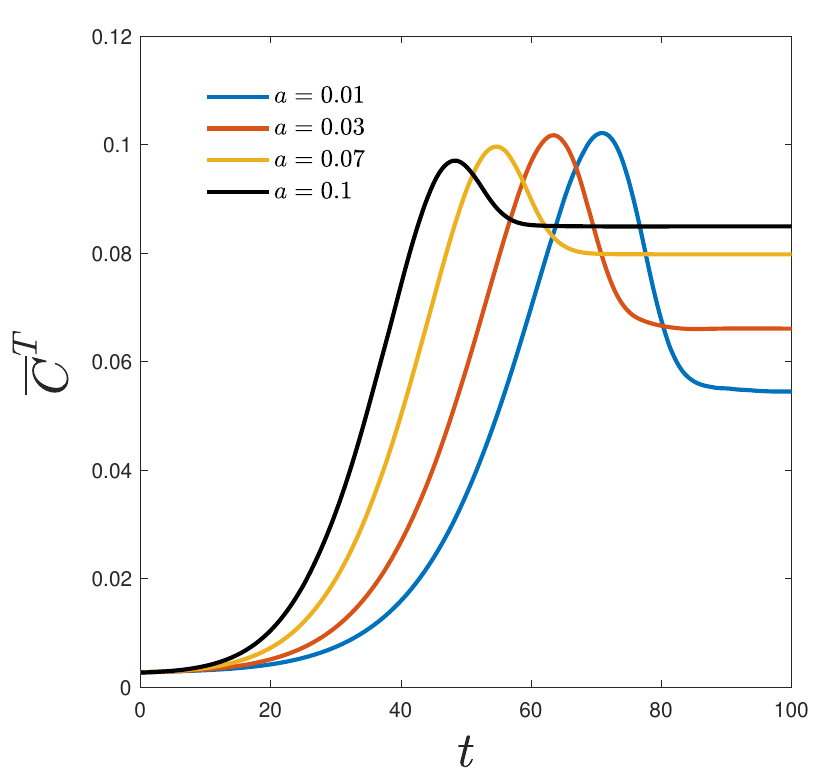}
\caption{}
\label{fig:6b}
\end{subfigure}
\begin{subfigure}[b]{0.5\textwidth}
\centering
\includegraphics[width=\textwidth]{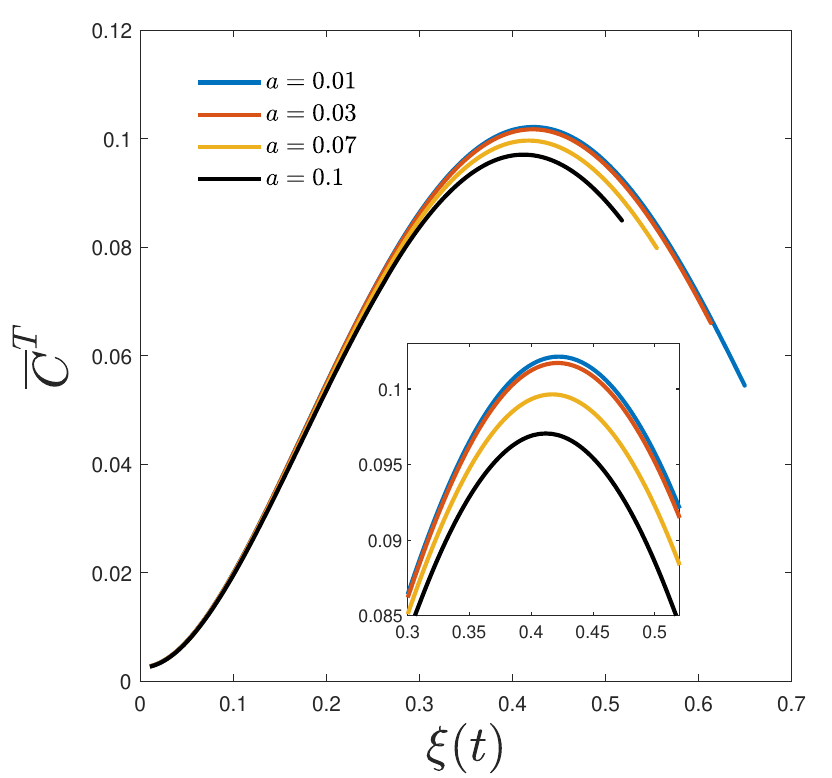}
\caption{}
\label{fig:6c}
\end{subfigure}
\caption{Variation of (a) cardiac troponin (cTn) concentration (b) average cardiac troponin ($\overline{\textit{cTn}}$) concentration with time (c) average cardiac troponin ($\overline{\textit{cTn}}$) concentration with $\xi(t)$ corresponding to $a=1 \times10^{-2},~3\times10^{-2},~ 7\times10^{-2}$,~and $1 \times10^{-1}$.}
\end{figure}
\noindent
With increasing \textit{MPD}, narrowing down of the lumen is resulted. Hence, the number of RBCs that escape through the narrower lumen is expected to reduce with an increased magnitude of \textit{MPD}. In this regard, Figures \ref{fig:6a}-\ref{fig:6b} respectively analyze the changes in local \cTn~concentration over time at a specific point within the AC area and the $\overline{\textit{cTn}}$ concentrations within the lumen of the artery affected by atherosclerosis, respectively, considering different radii $a=1 \times10^{-2},~3\times10^{-2},~ 7\times10^{-2}$,~and $1 \times10^{-1}$ of the plug region. Local \cTn~is found to behave similar to \textit{MPD} with time corresponding to the various $a$. Figure \ref{fig:6a} demonstrates the acceleration of \textit{cTn} concentration over time as the \textit{PRR} value increases. For larger $a$, \cTn~concentration attains a stable value more rapidly than for smaller $a$. Furthermore, as the \textit{PRR} value increases, there is a decrease in the maximum local \textit{cTn}. Basically, \textit{MPD} increases over time and a subsequent rise in the \textit{PRR} value results in the lumen clearance (\textit{CL}) reduction. As a result, the motion of the red blood cells gets resisted, which may lead to oxygen deprivation in the heart muscle. As the Figure \ref{fig:6a} is concerned, the situation of oxygen deprivation would be more significant when the pellet sizes are smaller (since \textit{MPD} attends higher magnitudes). However, the \textit{CL} becomes higher corresponding to the smaller \textit{PRR}. Therefore, one would expect higher oxygen possession in case of larger \textit{CL}. Apparently, this behavior is supposed to be an anomalous. But, it has been already discussed that the clearance has both the beneficial and detrimental aspects. On the other hand, Figure \ref{fig:6b} shows that the maximum $\overline{\textit{cTn}}$ values are nearly the same for all \textit{PRR}. However, there is a noticeable shift of the peak towards the right, indicating that the time to reach the maximum value is delayed as the pellet size decreases.  After reaching the peak, the $\overline{\textit{cTn}}$ profile decreases over time until it reaches a static value. Therefore, the effect of \cTn~will only last for a short period of time. When the state of oxygen deprivation to the heart muscles comes under control, the \cTn~concentration will drop to the average level. However, an emergency medical intervention is required to treat the issue that is depleting the oxygen supply to the heart muscle or brain tissue.\\

\noindent
The direct impact of \textit{MPD} on the $\overline{\textit{cTn}}$ concentration varies from the above observations. In this context, Figure \ref{fig:6c} shows that when the \textit{MPD} is less than or equal to $0.3$, the effect of the \textit{PRR} values on the $\overline{\textit{cTn}}$ is minimal. However, beyond this value, the relatively significant impact of \textit{PRR} on the combined behavior of \textit{MPD} and $\overline{\textit{cTn}}$ becomes evident. A comparatively significant variation is observed with smaller \textit{PRR}. Specifically, for $a=0.1$ and $a=0.01$, we can identify the respective maximum values of $\overline{\textit{cTn}}$ for $\xi(t)\approx0.43$. There is a higher likelihood of plaque development in the case of $a=0.01$ compared to $a=0.1$. In the former case, due to a wider lumen, the growing plaque experiences reduced normal stresses in the form of resistances compared to the latter. Consequently, a higher $\overline{\textit{cTn}}$ corresponds to the plaque growth process when the \textit{PRR} value is smaller. These observations clearly indicate an inverse relationship between $a$ and \textit{cTn}. In addition, such an inverse relation exits between \textit{MPD} and \textit{CL}. Therefore, we must concentrate upon the behaviorial relationship exists between $a$ and \textit{CL}.\\

\begin{figure}[htp]
\centering
\includegraphics[width=0.5\linewidth]{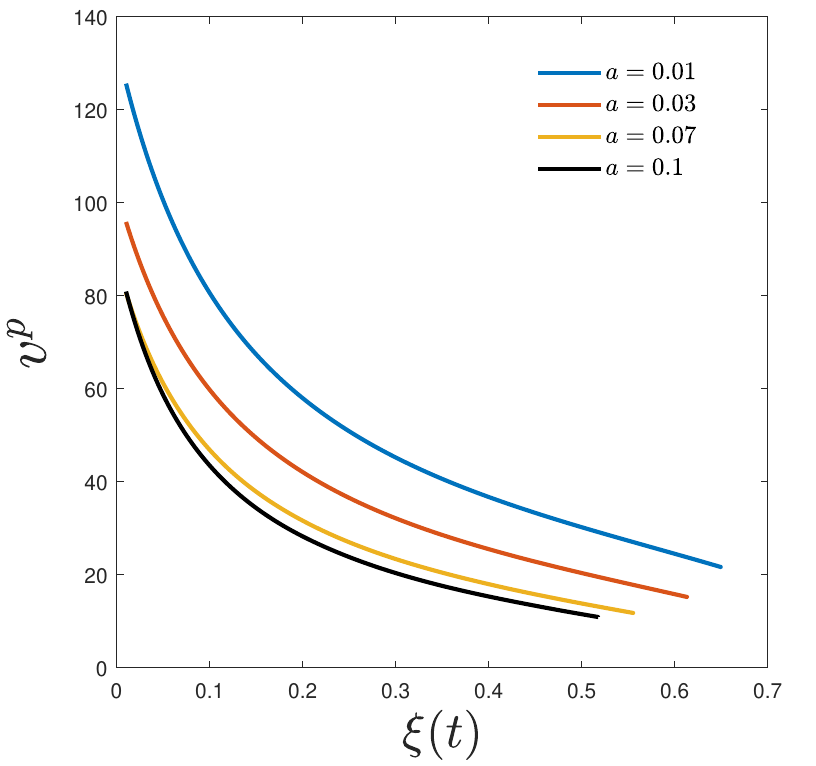}
\caption{Variation of plug flow velocity of pellets with the \textit{MPD} for various $a=1 \times10^{-2},~3\times10^{-2},~ 7\times10^{-2}$,~and $1 \times10^{-1}$.}
\label{fig:11}
\end{figure}
\noindent
We may discuss the issue of oxygen deprivation in term of the variation of velocity of the pellet stream $v_{p}$ with \textit{MPD}. In this context, the changes in $v_{p}$ over the \textit{MPD} is described in Figure \ref{fig:11}, considering various values of \textit{PRR} considered above. We observe a declining trend in the pellet velocity curve with increasing \textit{MPD}. Consequently, a reduction in clearance constrains the passage of blood cells through the artery, leading to a decrease in blood cell velocity. \textit{PRR} with smaller size experience a larger velocity drop with respect to the growth of $\xi(t)$ (\textit{MPD}). On the other hand, a significant small difference in the velocity drop can be observed between the cases $a=0.07$ and $a=0.1$. Therefore, above observation indicates a progressive decline in the velocity curve with increasing $a$. With a smaller blood cell radius, there is an increased chance of unobstructed passage through the artery. Hence, the above discussion indicates that \textit{cTn} levels in the blood increase with \textit{MPD} when the surrounding heart muscles experience oxygen deprivation and may release \textit{cTn} into the bloodstream. \\

\begin{figure}[htp]
\centering
\begin{subfigure}[b]{0.49\textwidth}
\centering
\includegraphics[width=\textwidth]{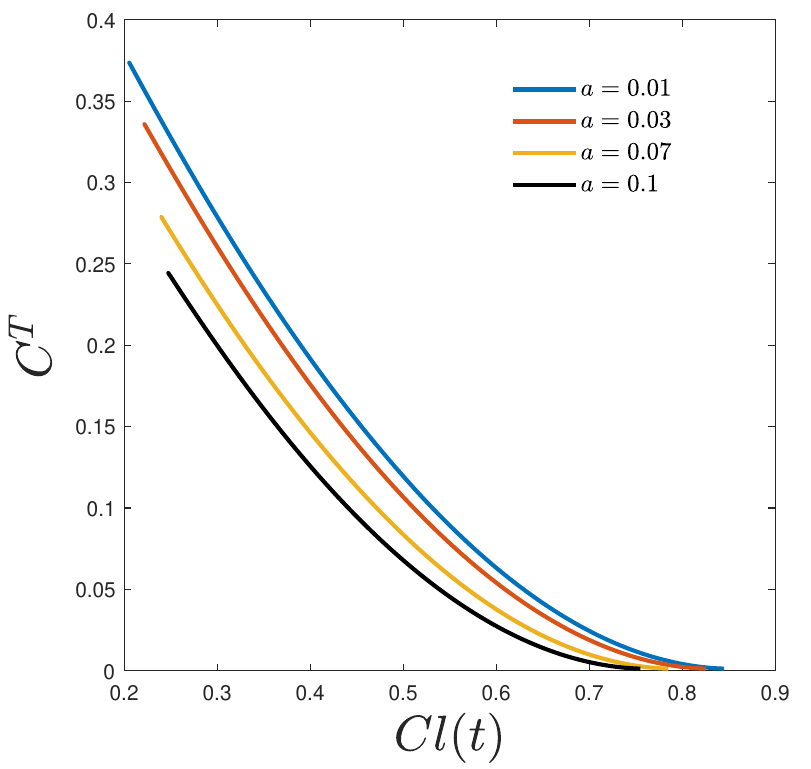}
\caption{}
\label{fig:7a}
\end{subfigure}
\begin{subfigure}[b]{0.49\textwidth}
\centering
\includegraphics[width=\textwidth]{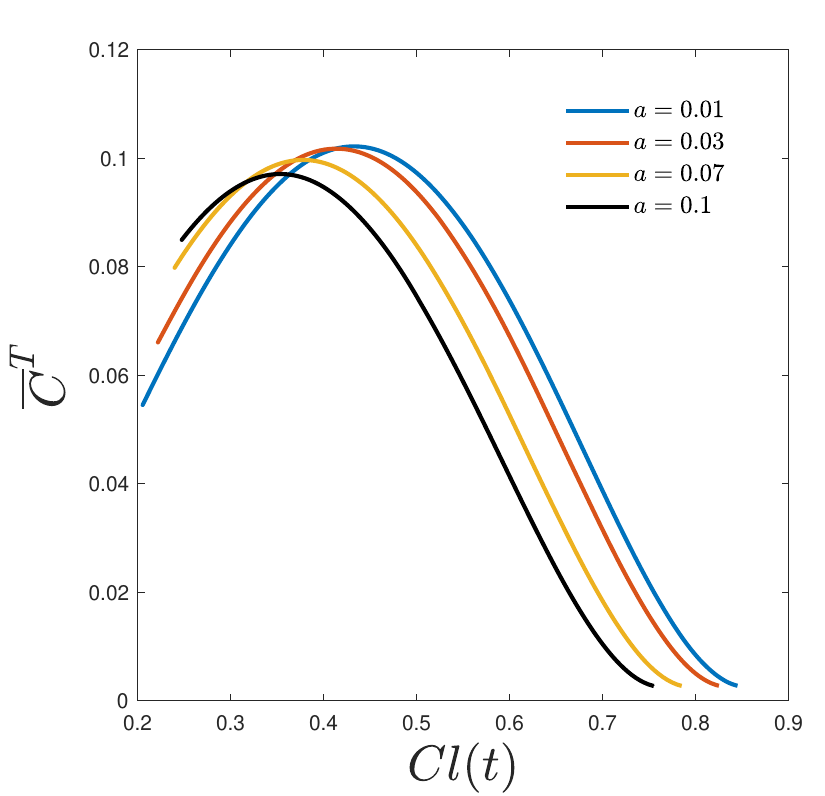}
\caption{}
\label{fig:7b}
\end{subfigure}
\caption{Variation of (a) cardiac troponin (cTn) and (b) average cardiac troponin ($\overline{\textit{cTn}}$) with respect to the clearance corresponding to $a=1 \times10^{-2},~3\times10^{-2},~ 7\times10^{-2}$,~and $1 \times10^{-1}$.}
\end{figure}
\noindent
The changes in \textit{cTn} and $\overline{\textit{cTn}}$ in blood with \textit{CL} are portrayed in figures \ref{fig:7a}-\ref{fig:7b}, considering various magnitudes of $(a)$ mentioned before. The \textit{cTn} concentrations initially decline gradually with the \textit{CL} and eventually reach near zero. This incident occurs because higher \textit{CL} levels indicate an unobstructed artery with no plaque, where \textit{cTn} concentration approaches zero, as observed in a healthy artery. As Figure \ref{fig:7a} is concerned, oxygen deprivation would be more significant when the \textit{PRR} value is smaller—however, the clearance increases, corresponding to the smaller \textit{PRR} values. Therefore, one would expect higher oxygen possession in case of larger clearance. As discussed, this bizarre behavior depicts the detrimental aspect of clearance besides its beneficial aspect. On the other hand, in Figure \ref{fig:7b}, the $\overline{\textit{cTn}}$ is initially seen to rise concerning clearance, reaching a peak before being followed by a decline in the later stages and ultimately approaching near zero. $\overline{\textit{cTn}}$ attend the maximum corresponding to $a=0.01$ among the all other values considered. Subsequently, after reaching the peak, all the curves corresponding to the considered values of $a$ decrease with a declining rate to the ambient value. However, the highest value of $\overline{\textit{cTn}}$ can be achievable only when $a=0.01$ among all other values. Moreover, before attending the peak, one can find that the highest increment rate can be observed corresponding to $a=0.01$ as compared to the other values $a$. Like the behavior of \textit{MPD} upon $\overline{\textit{cTn}}$, the internal geometry dependency of the atherosclerotic artery plays a role in this present context, which also explains the graphical nature between \textit{CL} and $\overline{\textit{cTn}}$.
\subsection{$\phi^{\alpha}$ Versus $\xi(t)$ (\textit{MPD})}
\begin{figure}[htp]
\centering
\includegraphics[width=0.5\linewidth]{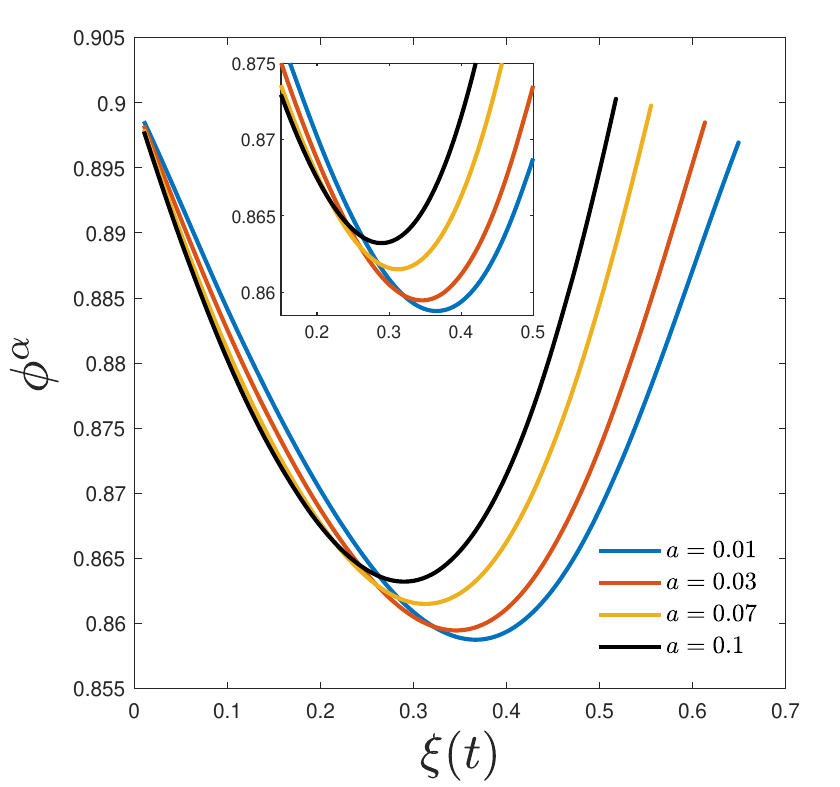}
\caption{Variation of inflammatory tissue volume within plaque with the \textit{MPD} for various $a=1 \times10^{-2},~3\times10^{-2},~ 7\times10^{-2}$,~and $1 \times10^{-1}$.}
\label{fig:12a}
\end{figure}

\noindent
The volume of inflammatory tissue $\varphi^{\alpha}$ initially decreases as \textit{MPD} increases, which suggests the non-inflammatory tissue initially contributes to plaque growth. The size of the pellet has a minimal impact on $\varphi^{\alpha}$ when $\xi(t)$ is less than or equal to $0.3$. However, beyond this value of \textit{MPD}, the parameter $a$ has a significant impact on $\varphi^{\alpha}$. Precisely, as the magnitude of \textit{MPD} exceeds $0.3$, the larger size of the pellet significantly impacts the inflammatory tissue volume. An increasing trend of $\varphi^{\alpha}$ is observed beyond $\xi(t)=0.3$, indicating the production of inflammatory tissues from the non-inflammatory counterpart. Apparently, one can not expect a direct impact of pellet size on the inflammatory tissue volume. The probable justification would be assuming a constant volumetric flow rate across the atherosclerotic artery. This assumption would suggest that the velocity of the pellet stream $v_{p}$ (velocity of plug flow of pellets) increases with the reduced \textit{PRR} value (see Figure \ref{fig:12a}). On the other hand, the constant volumetric flow rate assumption enables us to understand that $\varphi^{\alpha}$ is connected reciprocally with $v_{p}$ as $v^{\alpha}_{z}$ remains constant at any arterial cross-section. Therefore, the above arguments suggest that increasing pellet size positively influences the inflammatory tissue volume when the \textit{MPD} increases to a significant value, and it is $\xi(t)=0.3$ as obtained from this study.
\subsection{Qualitative Agreement with the Clinical Study}
\begin{figure}[h!]
\centering
\includegraphics[width=0.5\textwidth]{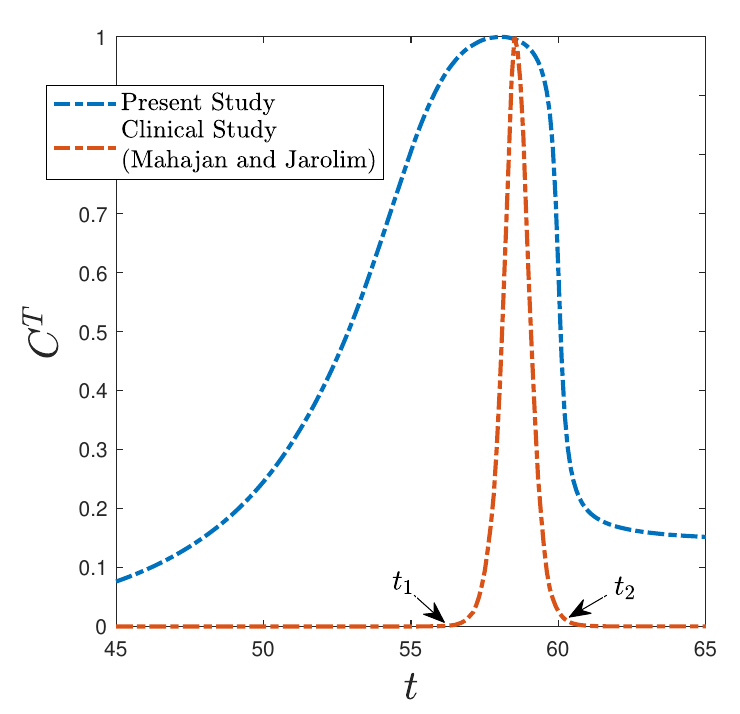}
\caption{Comparison between the present study and the clinical study done by \citet{mahajan2011interpret} regarding the temporal variation of cardiac troponin concentration.}
\label{fig:13a}
\end{figure}

\noindent
The present study simulates the plaque progression phenomenon within the dimensionless time interval $0$ to $100$. Each integer point within this interval is equivalent to $7$ days. Therefore, the present theoretical investigation can forecast the overall plaque growth process for over two years. Based on some clinical evidence, the present model considers that secretion of \cTn~and its corresponding concentration hike in the blood depends on the rise in plaque depth. This situation happens because the heart muscle realizes a gradual increase in oxygen deprivation due to plaque progression. The present study cannot predict the onset of any cardiovascular emergency. Moreover, it is difficult to predict through any survey at which position of plaque growth any cardiovascular emergency can arise unless one includes the phenomena called plaque rupturing and thrombus formation at the plaque built-up area. Hence, we manually set a time $t=t_{1}$ in the present investigation such that beyond which a gradual growth in local \cTn~is noticed with increased plaque depth. In our earlier graphical results, we set $t_{1}=40$ (equivalent to $280$ days). We introduce an additional term in the \cTn~generation term described in equation (\ref{equ108}), indicating that after a specific time $(t_2)$, the \cTn~concentration starts to decay. This adjustment assumes that specific medical interventions, such as surgery or medication, have already taken place, causing the concentration level to decrease in the bloodstream. We set $t_{2}=61$ (equivalent to $427$ days) in the present study. The modified \textit{cTn} generation term described in equation (\ref{equ108}) is represented as follows:
\begin{eqnarray}\label{equ115}
G(t,C^\TT)&=&\frac{\Lambda_{1}+\left[\left\{{\Lambda_2{\xi^{n}(t)}}/{(1+\Lambda_3{\xi^{n}(t)})}\right\}\mathcal{H}(t-t_1)\right] C^\TT}{1+\left[\left\{{\Lambda_4\xi^{n}(t)}/{(1+\Lambda_{5}{\xi^{n}(t)})}\right\}\mathcal{H}(t-t_1)\right]C^\TT}-\Lambda_6 C^\TT \mathcal{H}(t-t_1)\\\nonumber
&-&\left[\frac{\Lambda_{12}+\Lambda_{10} \xi^{n}(t)C^\TT}{1+\Lambda_{11}\xi^{n}(t)}\right]\mathcal{H}(t-t_2),
\end{eqnarray}
\noindent
We set the parameter $\Lambda_{11}$ to $0.01$. After certain medical interventions, \textit{cTn} level starts to decay in the bloodstream. In this context, the parameters $\Lambda_{10}$ and $\Lambda_{12}$ in Eq. (\ref{equ115}) are assigned with values $1$ and $20$ respectively. The solution to the diffusion equation of \textit{cTn}, considering the modified generation term and the above assumption, can be expressed as:
\begin{equation}\label{equ116}
C^\TT(r;r_{0},a)=\mathcal{C}_3J_0(q_1r)+\mathcal{C}_4Y_0(q_1r)+\frac{{\Lambda_{12}H(t-t_1)}-\Lambda_{1}}{{q^2_1}},
\end{equation}
where
\begin{equation*}
\mathcal{C}_4=\frac{\left(q_1^2\Lambda_{7} -({{\Lambda_{12}H(t-t_1)}-\Lambda_{1}})+q_1^2p_2^2\right)J_1(q_1a)}{q_1^2 \left[J_1(q_1a)Y_0(q_1r_0)-Y_1(q_1a)J_0(q_1r_0)\right]},
\end{equation*}
\begin{equation*}
\mathcal{C}_3=-\mathcal{C}_4 \frac{Y_1(q_1a)}{J_1(q_1a)},
\end{equation*}
\noindent
and
\begin{equation*}
q^2_1=\left|p^2_1-p^2_3\right|.
\end{equation*}
Note that
\begin{equation*}
p^2_1= \left|\frac{\Lambda_2\xi^{n}(t)}{1+\Lambda_{3}\xi^{n}(t)}-\Lambda_6\right|\mathcal{H}(t-t_1)~~~\textrm{and}~~~p^2_3= \left|\frac{\Lambda_{10}\xi^{n}(t)}{1+\Lambda_{11}\xi^{n}(t)}\mathcal{H}(t-t_2)\right|.
\end{equation*}
\noindent
\noindent
The plaque growth trend obtained from the present study gets clinically correlated with \citet{mahajan2011interpret}, which provided information about \cTn~concentration in patients with acute coronary syndromes. Their study asses the condition of three patients who reported chest discomfort at the emergency of a health care centre with or without previous history of cardiac-related issues. \cTn~concentration data from this clinical study was extracted and normalized using the Digitizer toolbox of Origin 2019b before comparison with the \cTn~and $\overline{\cTn}~$ concentrations obtained in our present study. The normalization process aligned the domain of clinical data with that of our study through an appropriate transformation, allowing us to focus on qualitative correspondence rather than quantitative. The \cTn~source in the present study corresponds to Eqs. (\ref{equ108}) and (\ref{equ116}) becomes influenced by $t=t_{1}$ when $\xi(t)$ begins to ascend. Therefore, a significant plaque development starts beyond $t_{1}=40$ ($\sim280$ days). On the contrary, the data obtained from \citet{mahajan2011interpret} corresponds to the patients already developing cardiac complications. The entire issue related to \cTn~becomes settled within $11$ days at most. Accordingly, \cTn~related data reported by \citet{mahajan2011interpret} is plotted between $t=t_{1}=56$ and $t=t_{2}=61$ in Figure \ref{fig:13a} to see a significant distinction in \cTn~profiles corresponding to both the studies. We are in a good position to compare.

\noindent
Figure \ref{fig:13a} illustrates this comparative analysis. This graph plots the \cTn~concentration data obtained from this study and \cTn~concentration data from the study of \citet{mahajan2011interpret} over time. The blue and red lines represent the \cTn~ concentration data received from both the present and clinical study. We observed a robust qualitative agreement between the clinical and our present study's data, indicating that the \cTn~concentrations from both studies exhibit similar behavior. Based on the observations of various factors, we can conclude that the concentration curves (obtained from both studies) display significantly elevated concentrations when plaque size is extensive, signifying conditions near the onset of heart attack or stroke. Cardiac troponin levels demonstrate a decreasing pattern after reaching a peak, following medical interventions such as medication, angioplasty, bypass surgery etc.
\section{Summary}
\noindent
Based on the biphasic mixture theory, we develop a mathematical model to analyze the dynamic evolution of atherosclerotic plaque in a cardiac artery and its simultaneous narrowing. This study investigates the evolution of cardiac troponin (\cTn~a biomarker), which is directly connected to the narrowing of a cardiac artery, referred to as arterial clearance. Our findings show an initial phase of exponential growth, followed by a stable state. Besides, we observe an initial phase of exponential decay, followed by a steady value where the arterial clearance converges. This study highlights the aiding and opposing impacts of arterial clearance on atherosclerosis. Furthermore, the velocity curve of pellets (blood cells) demonstrates a declining trend with increasing plaque depth, resulting in a reduced rate of blood cell passage in the artery. This relationship suggests that \textit{cTn} levels increase with plaque depth when surrounding heart muscles experience oxygen deprivation, releasing \textit{cTn} into the bloodstream. This study highlights a positive correlation between plaque depth and \cTn~concentration in the blood and a negative relationship between \cTn~ concentration and arterial clearance in atherosclerotic arteries. The comparison shows that the troponin concentration curve from this study behaves similarly to that of clinical research. The \cTn~concentration curve may effectively predict the early stages of myocardial infarction. The results of this study help us to identify the early myocardial infarction and cerebrovascular trouble related to atherosclerosis by monitoring \cTn~concentrations through some simple blood tests and subsequently by calculating \textit{MPD} (utilizing the \textit{cTn} data) to estimate the likelihood of myocardial infarction.
\section*{{ACKNOWLEDGMENTS}}
\noindent
The First author is thankful to the \textbf{Savitribai Jyotirao Phule Fellowship For Single Girl Child} (Award letter reference number: 82-7/2022(SA-III), Registration ID: UGCES-22-GE-WES-F-SJSGC-2025) of University Grant Commission, Govt. of India, for providing necessary financial support in order to conduct this work. Both the authors are thankful to Prof. G. P. Raja Sekhar, Professor of Mathematics, Indian Institute of Technology Kharagpur for his for valuable comments regarding lumenal blood flow and discussions on boundary conditions.
\section*{{AUTHOR DECLARATIONS}}
\noindent
\subsection*{Conflict of Interest}
The authors have no conflicts to disclose.
\section*{{DATA AVAILABILITY}}
\noindent
The data that support the findings of this study are available from the corresponding author upon reasonable request.
\appendix
\section{The Leading-Order Problem}\label{appendixA}
\noindent
For the leading-order problem, the governing equations are simplified to the following form:
\begin{equation}\label{equ45}
\frac{\partial P_0}{\partial r}=0,
\end{equation}
\begin{equation}\label{equ46}
-\frac{\partial P_0}{\partial z}+ \gamma^2 \frac{1}{r} \frac{\partial}{\partial r}\left(r \frac{\partial {(v^\alpha_z)}_0}{\partial r}\right)=0,
\end{equation}
\begin{equation}\label{equ49}
\left(\frac{1}{r}+\frac{\partial}{\partial r}\right)\left(\phi^\alpha_0 {(v^\alpha_r)}_0\right)+\frac{\partial}{\partial z}\left(\phi^\alpha_0 {(v^\alpha_z)}_0\right)=0,
\vspace{0.5cm}
\end{equation}
\begin{equation}\label{equ50}
r\frac{\partial}{\partial r}\left(\frac{1}{r}\frac{\partial}{\partial r}\left(r\frac{\partial}{\partial r}\left(\frac{1}{r}\frac{\partial \psi_0}{\partial r}\right)\right)\right)=0.
\end{equation}
The corresponding boundary conditions are
\begin{enumerate}[label=(\roman*)]
\item On $r=1$,
\begin{equation}\label{equ51}
{(v^\alpha_z)}_0=0~~ \text{and}~~ {(v^\alpha_r)}_0=0.
\end{equation}
\item For some $z_0\in[-1, 1]$, at $r=r_0$,
\begin{equation}\label{equ52}
\phi^\alpha_0 {(v^\alpha_z)}_0={(v^l_z)}_0~,~~{(v^l_r)}_0=0~,
\end{equation}
\begin{equation}\label{equ53}
(P_0-P_0^l)=0~~\text{and}~~\gamma^2\frac{\partial {(v^\alpha_z)}_0}{\partial r}= \beta^2 \frac{\partial {(v^l_z)}_0}{\partial r}.
\end{equation}
\item At $r=a$
\begin{equation}\label{equ54}
{(v^l_z)}_0=v^p~~,~~{(v^l_r)}_0=0~~\text{and}~~P_0^l=0.
\end{equation}
\item Flux Condition:
\begin{equation}\label{equ55}
\int_0^{a} v^p\,r\,dr+\int_{a}^{B(z,t)} {(v^l_z)}_0\,r\,dr+\int^1_{B(z,t)}\phi^\alpha_0 {(v^\alpha_z)}_0 \,r\,dr=1.
\end{equation}
\end{enumerate} 
\section{The Leading-Order Solution}\label{appendixB}
\noindent
The solution of Eq. (\ref{equ46}) with the help of Eq. (\ref{equ45}) is given by
\begin{equation}\label{equ56}
{(v^\alpha_z)}_0= \frac{r^2}{4\gamma^2}\frac{d P_0}{d z}+a_1(z) \log\left(\frac{r}{r_0}\right)+a_2(z),
\end{equation}
 \noindent
 By utilizing the continuity equation (\ref{equ49}) and making the assumption that the leading-order term of the volume fraction expansion i.e. $\phi^i_0$, is constant, one can derive the following solution
 \begin{equation}\label{equ57}
   {(v^\alpha_r)}_0= \frac{ (1-r^4)}{16r \gamma^2}\frac{d ^2 P_0}{d z^2}-\frac{1}{2}a'_1(z)r \log\left(\frac{r}{r_0}\right)-\frac{ (1-r^2)}{4r}(a'_1(z)-2a'_2(z))+\frac{a_3(z)}{r},
 \end{equation}
 The solution of Eq. (\ref{equ50}) is given by
 \begin{equation}\label{equ58}
   \psi_{0}= c_1(z)(r^4-a^4)+ c_2(z) (r^2) \log \left( \frac{r}{r_0} \right)+c_3(z) (r^2-a^2)+c_4(z),
 \end{equation}
 By using this solution, we can determine the leading order velocities of the fluid within the lumen
 \begin{equation}\label{equ59}
     {(v^l_r)}_0=-\frac{1}{r}\frac{\partial \psi_0}{\partial z}= -\frac{c'_1(z)(r^4-a^4)}{r}- c'_2(z)r \log \left(\frac{r}{r_0}\right)-\frac{c'_3(z)(r^2-a^2)}{r} - \frac{c'_4(z)}{r},
 \end{equation}
 \noindent
 and
\begin{equation}\label{equ60}
     {(v^l_z)}_0=\frac{1}{r}\frac{\partial \psi_0}{\partial r}= 4 r^2 c_1(z)+ c_2(z) \left( 2 \log \left( \frac{r}{r_0}\right)+1 \right)+2 c_3(z).
 \end{equation}
 \noindent
 By applying the boundary conditions, we can derive the expressions of the functions $ a'_1(z),a'_2(z),a_3(z),\\c'_1(z),c'_2(z),c'_3(z),c'_4(z) $
 \begin{equation}\label{equ61}
   a'_1(z)=D(r_0)\frac{d^2 P_0}{dz^2},
 \end{equation}
 \begin{equation}\label{equ62}
   a'_2(z)= \left[D(r_0) \log(r_0)- \frac{1}{4\gamma^2}\right] \frac{d^2 P_0}{dz^2},
 \end{equation}
 \begin{equation}\label{equ63}
   a_3(z)=-\frac{D(r_0)}{2} \log(r_0)\frac{d^2 P_0}{dz^2},
 \end{equation}
 \begin{equation}\label{equ64}
   c'_1(z)=0,
 \end{equation}
 \begin{equation}\label{equ65}
   c'_2(z)=C(r_0)\frac{d^2 P_0}{dz^2},
 \end{equation}
 \begin{equation}\label{equ66}
   c'_3(z)=C(r_0)\frac{a^2 \log \left( \frac{a}{r_0}\right)}{r_0^2-a^2}\frac{d^2 P_0}{dz^2},
 \end{equation}
 \begin{equation}\label{equ67}
   c'_4(z)=-C(r_0)a^2 \log\left(\frac{a}{r_0}\right) \frac{d^2 P_0}{dz^2},
 \end{equation}
where
\begin{equation*}
  A(r_0)= \frac{2a^2\log\left( \frac{a}{r_0}\right)}{r_0^2-a^2}+ 2\log\left( \frac{a}{r_0}\right)+1,
  \vspace{0.1cm}
\end{equation*}
\begin{equation*}
  B(r_0)= \frac{A(r_0) \left( (1-r_0^2)/(4\gamma^2)+ (r_0^2 \log(r_0))/(2 \gamma^2)\right)} {\left((2 \beta^2 \phi^\alpha_{*})/\gamma^2\right)\log(r_0)-\left(2a^2\log\left( \frac{a}{r_0}\right)/(r_0^2-a^2)\right)-1},
  \vspace{0.1cm}
\end{equation*}
\begin{equation*}
  C(r_0)= \frac{B(r_0)}{A(r_0)},
  \vspace{0.3cm}
\end{equation*}
\begin{equation*}
  D(r_0)= \frac{2 \beta^2 \phi^\alpha_{*}}{\gamma^2} C(r_0)- \frac{r_0^2}{2 \gamma^2}.
  \vspace{0.1cm}
\end{equation*}
The constant volumetric flux condition helps us to evaluate the pressure gradient as follows:
\begin{equation}\label{equ68}
    \frac{\partial P_0}{\partial z}= \frac{1-\frac{v_pa^2}{2}}{K},
\end{equation}
where
\begin{equation*}
\begin{split}
     K=&C(r_0) \left( r \log \left( \frac{r}{r_0}\right)-a^2 \log \left( \frac{r}{r_0}\right) \right) +\frac{C(r_0)}{r_0-a^2}a^2(r^2-a^2)\log\left(\frac{a}{r_0}\right)\\
    & +\phi_{0}^\alpha\left[ \frac{1}{16 \gamma^2}(1-{r}^4)-\frac{D(r_0)}{2}{r}^2 \log\left(\frac{r}{r_0}\right)+\frac{1-{r}^2}{2}\left( D(r_0)\left(\log(r_0)-\frac{1}{2}\right)-\frac{1}{4 \gamma^2} \right)-\frac{D(r_0)}{2}\log(r_0)\right].
 \end{split}
\end{equation*}
\noindent

\section{The $\textit{O}(\delta^2)$ Problem}\label{appendixC}
\noindent
The governing equations corresponding to the first-order problem can be expressed as
\begin{equation}\label{equ69}
-\phi^\alpha_0\frac{\partial P_1}{\partial r}- \phi^\alpha_1 \frac{\partial P_0}{\partial r}+\frac{\gamma^2}{3}\left[\frac{\partial}{\partial r}\left\{\frac{1}{r}\frac{\partial}{\partial r}\left(r{(v^\alpha_r)}_0\right)+\frac{\partial {(v^\alpha_z)_0}}{\partial z}\right\}\right]+ \gamma^2 \left[\frac{1}{r}\frac{\partial}{\partial r}\left(r\frac{\partial {(v^\alpha_r)}_0}{\partial r}\right)-\frac{{(v^\alpha_r)}_0}{r^2}\right]- {(v^\alpha_r)}_0=0,
\end{equation}
\begin{equation}\label{equ70}
-\phi^\alpha_0\frac{\partial P_1}{\partial z}- \phi^\alpha_1\frac{\partial P_0}{\partial z}+\frac{\gamma^2}{3}\left[\frac{\partial}{\partial z}\left\{\frac{1}{r}\frac{\partial}{\partial r}\left(r{(v^\alpha_r)}_0\right)+\frac{\partial {(v^\alpha_z)}_0}{\partial z}\right\}\right]+\gamma^2 \left[\frac{1}{r}\frac{\partial}{\partial r}\left(r\frac{\partial {(v^\alpha_z)}_1}{\partial r}\right)+\frac{\partial^2 {(v^\alpha_z)}_0}{\partial z^2}\right]-{(v^\alpha_z)}_1=0,
\end{equation}
\begin{equation}\label{equ71}
- \phi^\beta_0\frac{\partial P_1}{\partial r}- \phi^\beta_1\frac{\partial P_0}{\partial r}+ {(v^\alpha_r)}_0=0,
\end{equation}
\begin{equation}\label{equ72}
- \phi^\beta_0\frac{\partial P_1}{\partial z}- \phi^\beta_1\frac{\partial P_0}{\partial z}+ {(v^\alpha_z)}_1=0,
\end{equation}
\begin{equation}\label{equ73}
\left(\frac{1}{r}+\frac{\partial}{\partial r}\right)\left(\phi^\alpha_0 {(v^\alpha_r)}_1+\phi^\alpha_1 {(v^\alpha_r)}_0\right)+\frac{\partial}{\partial z}\left(\phi^\alpha_0 {(v^\alpha_z)}_1+\phi^\alpha_1 {(v^\alpha_z)}_0\right)=0,
\end{equation}
\begin{equation}\label{equ74}
r\frac{\partial}{\partial r}\left(\frac{1}{r}\frac{\partial}{\partial r}\left(r\frac{\partial}{\partial r}\left(\frac{1}{r}\frac{\partial \psi_1}{\partial r}\right)\right)\right)+  2 \frac{\partial^2}{\partial z^2}\left(r\frac{\partial}{\partial r}\left(\frac{1}{r}\frac{\partial \psi_0}{\partial r}\right)\right)=0.
\vspace{0.5cm}
\end{equation}
\noindent
The corresponding boundary conditions are
\begin{enumerate}[label=(\roman*)]
\item at $r=1$,
\begin{equation}\label{equ75}
{(v^\alpha_z)}_1=0~~ \text{and}~~ {(v^\alpha_r)}_1=0.
\end{equation}
\item at $z=z_0\in[-1, 1]$ ~\text{and}~$r=r_0=B(z_{0},t)$,
\begin{equation}\label{equ76}
\phi^{\alpha}_1{(v^\alpha_z)}_1={(v^l_z)}_1~,~~{(v^l_r)}_1=0~,
\end{equation}
\begin{equation}\label{equ77}
-(P_1-P_1^l)-\frac{2\gamma^2}{3} \left( \frac{1}{r} \frac{\partial}{\partial r}(r{(v^\alpha_r)}_0)+ \frac{\partial {(v^\alpha_z)}_0}{\partial z}\right)+ 2 \left(\gamma^2\frac{\partial {(v^\alpha_r)}_0}{\partial r}- \beta^2 \frac{\partial {(v^l_r)}_0}{\partial r}\right) =0,
\end{equation}
\begin{equation}\label{equ78}
\gamma^2 \left( \frac{\partial {(v^\alpha_r)}_0}{\partial z}+\frac{\partial {(v^\alpha_z)}_1}{\partial r}\right)= \beta^2 \left( \frac{\partial {(v^l_r)}_0}{\partial z}+ \frac{\partial {(v^l_z)}_1}{\partial r}\right),
\end{equation}
\noindent
and
\begin{equation}\label{equ79}
-(P_1-P_1^l)-\frac{2\gamma^2}{3} \left( \frac{1}{r} \frac{\partial}{\partial r}(r{(v^\alpha_r)}_0)+ \frac{\partial {(v^\alpha_z)}_0}{\partial z}\right)+ 2 \left(\gamma^2\frac{\partial {(v^\alpha_z)}_0}{\partial z}- \beta^2 \frac{\partial {(v^l_z)}_0}{\partial z}\right) =0.
\end{equation}
\item at $r=a$
\begin{equation}\label{equ80}
{(v^l_z)}_1=0~~,~~{(v^l_r)}_1=0,
\end{equation}
\noindent
and
\noindent
\begin{equation}\label{equ81}
P_1^l+\beta^2 \left( \frac{\partial v^l_r}{\partial r}+\frac{\partial v^l_z}{\partial z}\right)=0.
\end{equation}
\item Flux Condition
\begin{equation}\label{equ82}
0= \int_{a}^{R(z,t)} {(v^l_z)}_1\,r\,dr+\int^1_{R(z,t)}\phi^\alpha_1 {(v^\alpha_z)}_1 \,r\,dr.
\end{equation}
\end{enumerate}
\noindent
Hence the general solution of the plaque velocity in the r-direction is determined up to $\textit{O}(\delta^2)$ as
\begin{equation}\label{equ97}
v^\alpha_r = {(v^\alpha_r)}_0 + {\delta^2} {(v^\alpha_r)}_1+ \textit{O}(\delta^4).
\end{equation}
\section{$\textit{O}(\delta^2)$ Solution}\label{appendixD}
\noindent
By solving the Eqs. (\ref{equ70}) and (\ref{equ72}), the following results can be obtained:
 \begin{equation}\label{equ83}
  \begin{aligned}
   {(v^\alpha_z)}_1 = & -\frac{r^2(r^2-2)}{32\gamma^2}  \frac{d ^3P_0}{d z^3}+ \frac{a''_1(z)}{2}(1-r^2)\log \left( \frac{r}{r_0}\right)-\frac{(1-r^2)}{4}\left(\frac{a'_4(z)}{\gamma^2}-a''_2(z)+2a''_1(z)\right)\\
   & + a_5 \log \left( \frac{r}{r_0}\right)+a_6(z),
   \end{aligned}
 \end{equation}
 \noindent
 and
 \begin{equation}\label{equ84}
  \begin{aligned}
   \phi_1^\alpha  = & -\frac{1}{\frac{d P_0}{d z}} \left[ \frac{d ^3 P_0}{d z^3} \left\{ -\frac{(1-r^2)}{4}\phi_0^\beta + \frac{r^2}{32 \gamma^2}(2-r^2)\right\}+ a''_1(z) \log \left(\frac{r}{r_0}\right)\left\{\phi_0^\beta\gamma^2+\frac{(1-r^2)}{2}\right\}\right]\\
   & -\frac{1}{\frac{d P_0}{d z}} \left[- \frac{(1-r^2)}{4}\left(\frac{a'_4(z)}{\gamma^2}-a''_2(z)+2a''_1(z)\right)+a_5(z)\log \left( \frac{r}{r_0}\right)+ (a_6(z)- \phi_0^\beta a'_4(z))\right],
  \end{aligned}
 \end{equation}
 \noindent
 Using the continuity equation (\ref{equ74}), one can obtain
 \begin{equation}\label{equ85}
   \begin{split}
   {(v^\alpha_r)}_1=&- \frac{\phi_1^\alpha}{\phi_0^\alpha} {(v^\alpha_r)}_0- \frac{\partial^4 P_0}{\partial z^4} \left[ \frac{1}{32\gamma^2} \frac{(1-r^6)}{6r} - a(r_0)\frac{(1-r^4)}{4r}- c(r_0)\frac{(1-r^2)}{2r} + \log\left(\frac{r}{r_0}\right)\left(b(r_0)\frac{r}{2}-D(r_0)\frac{r^3}{8}\right)\right]\\
   &-\frac{\partial^4 P_0}{\partial z^4}\frac{j^*(r_0)}{r\phi^\alpha_0}-\frac{1}{\phi_0^\alpha} \frac{\partial^4 P_0}{\partial z^4}\left[-\frac{1}{128\gamma^4} \frac{(1-r^8)}{8r}-m(r_0)\frac{(1-r^6)}{6r}-p(r_0)\frac{(1-r^4)}{r}-s(r_0)\frac{(1-r^4)}{r}\right]\\
  & -\frac{1}{\phi_0^\alpha} \frac{\partial^4 P_0}{\partial z^4} \left[\log\left(\frac{r}{r_0}\right) \left( \frac{g(r_0)r^5}{6}+o(r_0)r^3+q(r_0)r\right) +\left(\log\left(\frac{r}{r_0}\right)\right)^2 \left( n(r_0)r^3-D(r_0)e(r_0)\frac{r}{2}\right)\right].
  \end{split}
  \end{equation}
  \noindent
  where
  \begin{equation*}
    a(r_0)=\frac{\alpha^2}{8}+\frac{\alpha^2}{4}i^*(r_0)+\frac{D(r_0)}{4}\left(\frac{5}{2}-\log(r_0)\right),
\end{equation*}
\begin{equation*}
    b(r_0)=\frac{D(r_0)}{2}+g^*(r_0)i^*(r_0)+h^*(r_0),
\end{equation*}
\begin{equation*}
    c(r_0)=\frac{D(r_0)}{4}(\log(r_0)-3)-\frac{3\alpha^2}{32}-\frac{\alpha^2}{4}i^*(r_0)+\frac{1}{2}(2\log(r_0)-1)(g^*(r_0)i^*(r_0)+h^*(r_0)),
\end{equation*}
\begin{equation*}
    d(r_0)= \frac{\phi^\beta_0}{4}+\frac{\alpha^2}{8}+\frac{\alpha^2}{4}i^*(r_0)+\frac{D(r_0)}{4}(2-\log(r_0)),
\end{equation*}
\begin{equation*}
    e(r_0)=D(r_0)\left( \frac{\phi^\beta_0}{\alpha^2}+\frac{1}{2}\right)+g^*(r_0)i^*(r_0)+h^*(r_0),
\end{equation*}
\begin{equation*}
    f(r_0)=-\frac{\phi^\beta_0}{4}-\frac{3\alpha^2}{32}-i^*(r_0)\left( \frac{\alpha^2}{4}+\phi^\beta_0 \right)+\frac{D(r_0)}{4}(\log(r_0)-2)+(g^*(r_0)i^*(r_0)+h^*(r_0))\log(r_0),
\end{equation*}
\begin{equation*}
    g(r_0)=\frac{5\alpha^2}{32}D(r_0)
\end{equation*}
\begin{equation*}
    h(r_0)=-\frac{\alpha^2}{4}d(r_0)+\frac{\alpha^2}{32}\left( D(r_0)\log(r_0)-\frac{\alpha^2}{4} \right),
\end{equation*}
\begin{equation*}
    i(r_0)=-\frac{\alpha^2}{4}e(r_0)-D(r_0)d(r_0)+\frac{1}{2}D(r_0)\left( D(r_0)\log(r_0)-\frac{\alpha^2}{4} \right),
\end{equation*}
\begin{equation*}
    j(r_0)=-\frac{\alpha^2}{4}f(r_0)-d(r_0)\left(D(r_0)\log(r_0)-\frac{\alpha^2}{4}\right),
\end{equation*}
\begin{equation*}
    k(r_0)=-D(r_0)f(r_0)-\left( D(r_0)\log(r_0)-\frac{\alpha^2}{4} \right)e(r_0),
\end{equation*}
\begin{equation*}
    m(r_0)=h(r_0)-\frac{g_(r_0)}{6},
\end{equation*}
\begin{equation*}
    n(r_0)=\frac{1}{8}(D(r_0))^2,
\end{equation*}
\begin{equation*}
    o(r_0)=\frac{i(r_0)}{4}-\frac{1}{16}(D(r_0))^2,
\end{equation*}
\begin{equation*}
    p(r_0)=\frac{1}{4}j(r_0)-\frac{1}{16}i(r_0)+\frac{1}{64}(D(r_0))^2,
\end{equation*}
\begin{equation*}
    q(r_0)=e(r_0)D(r_0)+k(r_0),
\end{equation*}
\begin{equation*}
    s(r_0)=-\frac{1}{4}e(r_0)D(r_0)-\frac{k(r_0)}{4}-\frac{f(r_0)}{2}\left( D(r_0)\log(r_0)-\frac{\alpha^2}{4} \right).
\end{equation*}
\noindent
 The solution of equation (\ref{equ74}) is given by
 \begin{equation*}
  \begin{aligned}
   \psi_1 =&-\frac{c''_1(z)}{12}(r^6-a^6)-\frac{c''_2(z)}{4} r^4 \log \left( \frac{r}{r_0}\right)+ \frac{c_5(z)}{4}(r^4-a^4)\\
   &+c_6(z)r^2\log \left( \frac{r}{r_0}\right)+c_7(z)(r^2-a^2)+c_8(z),
   \end{aligned}
 \end{equation*}
 By using this solution, we can determine the first-order velocities of the fluid within the lumen
 \begin{equation}\label{equ87}
  \begin{aligned}
    {(v^l_r)}_1=&-\frac{1}{r}\frac{\partial \psi_1}{\partial z}\\
    =&\frac{c'''_1(z)}{12} \frac{(r^6-a^6)}{r} - \frac{c'''_2(z)}{4}r^3 \log \left( \frac{r}{r_0}\right)- \frac{c'_5(z)}{4}\frac{(r^4-a^4)}{r}\\
    & -c'_6(z)r \log \left( \frac{r}{r_0}\right)-c'_7(z)\frac{(r^2-a^2)}{r}- \frac{c'_8(z)}{r},
   \end{aligned}
 \end{equation}
 \begin{equation}\label{equ88}
  \begin{aligned}
   {(v^l_z)}_1= &\frac{1}{r}\frac{\partial \psi_1}{\partial r}\\
   = &-\frac{c''_1(z)}{2}r^4-\frac{c''_2(z)}{4}\left( r^2+ 4 r^2 \log \left( \frac{r}{r_0}\right)\right)+c_5(z)r^2\\
     &+ c_6(z)\left( 2 \log \left( \frac{r}{r_0}\right)+1 \right)+2 c_7(z).
  \end{aligned}
 \end{equation}
 By applying the boundary conditions, we can derive the expressions of the functions $ a'_4(z),a_5(z),a_6(z),\\d_1(z),d_2(z),d_3(z),d_4(z) $
 \begin{equation}\label{equ89}
     a'_4(z)=i^*(r_0) \frac{\partial^3 P_0}{\partial z^3},
 \end{equation}
 \begin{equation}\label{equ90}
     a_5(z)= \left[g^*(r_0)i^*(r_0)+h^*(r_0)\right]\frac{\partial^3 P_0}{\partial z^3},
 \end{equation}
 \begin{equation}\label{equ91}
     a_6(z)= a_5(z) \log(r_0)- \frac{1}{32 \gamma^2}\frac{\partial^3 P_0}{\partial z^3},
 \end{equation}
 \begin{equation}\label{equ92}
     a_7(z)=j^*(r_0) \frac{\partial^3 P_0}{\partial z^3},
 \end{equation}
 \begin{equation}\label{equ93}
     c_5(z)= \left( \frac{i^*(r_0)}{4 \beta^2}-\frac{1}{4}\left( F(r_0)+\frac{L(r_0)}{\beta^2}\right) \right) \frac{\partial^3 P_0}{\partial z^3},
 \end{equation}
 \begin{equation}\label{equ94}
     c_6(z)= \left[a^*(r_0)i^*(r_0)+b^*(r_0)\right]\frac{\partial^3 P_0}{\partial z^3},
 \end{equation}
 \begin{equation}\label{equ95}
     c_7(z)= \left[c^*(r_0)i^*(r_0)+d^*(r_0)\right]\frac{\partial^3 P_0}{\partial z^3},
 \end{equation}
 \begin{equation}\label{equ96}
     c_8(z)=\left[e^*(r_0)i^*(r_0)+f^*(r_0)\right]\frac{\partial^3 P_0}{\partial z^3}.
 \end{equation}
 \noindent
 where
 \begin{equation*}
    a^*(r_0)= \frac{(r_0^2-a^2)^2}{4 \beta^2 \left( 2(r_0^2-a^2)+4 r_0^2 \log\left( \frac{a}{r_0}\right)\right)},
\end{equation*}
\begin{equation*}
    b^*(r_0)= \frac{C(r_0)a^4 \log\left( \frac{a}{r_0}\right)+H(r_0)(r_0^2-a^2)-\frac{1}{4}\left( F(r_0)+\frac{L(r_0)}{\beta^2}\right)(r_0^2-a^2)^2}{2(r_0^2-a^2)+4 r_0^2 \log\left( \frac{a}{r_0}\right)},
\end{equation*}
\begin{equation*}
    c^*(r_0)= -\frac{1}{2}\left(\frac{a^2}{4 \beta^2}+a^*(r_0)\left( 2\log\left( \frac{a}{r_0}\right)+1 \right) \right),
\end{equation*}
\begin{equation*}
    d^*(r_0)= H(r_0)+\frac{a^2}{4}\left( F(r_0)+\frac{L(r_0)}{\beta^2}\right)- b^*(r_0)\left( 2\log\left( \frac{a}{r_0}\right)+1 \right),
\end{equation*}
\begin{equation*}
    e^*(r_0)= -a^2 \log\left(\frac{a}{r_0}\right) a^*(r_0),
\end{equation*}
\begin{equation*}
    f^*(r_0)= -a^2 \log\left(\frac{a}{r_0}\right) b^*(r_0)+ C(r_0) \frac{a^4}{4}\log\left(\frac{a}{r_0}\right),
\end{equation*}
\begin{equation*}
    g^*(r_0)= -\frac{r_0^2}{2 \gamma^2}+\frac{2 \phi^\alpha_{*} \beta^2}{\gamma^2}\left( \frac{r_0^2}{4 \beta^2}+a^*(r_0)\right),
\end{equation*}
\begin{equation*}
    h^*(r_0)= \frac{2 \phi^\alpha_{*} \beta^2}{\gamma^2} \left( -\frac{r_0^2}{4}\left( F(r_0)+\frac{L(r_0)}{\beta^2}\right)+b^*(r_0)\right)-M(r_0),
\end{equation*}
\begin{equation*}
    i^*(r_0)= \frac{J(r_0)+h^*(r_0)\log(r_0)+\frac{r_0^2}{4}\left( F(r_0)+\frac{L(r_0)}{\beta^2}\right)-b^*(r_0)-2d^*(r_0)} {\frac{(1-r_0^2)}{4 \gamma^2}- \log(r_0)g^*(r_0)+ \frac{r_0^2}{4 \beta^2}+ a^*(r_0)+2 c^*(r_0)},
\end{equation*}
\begin{equation*}
\begin{split}
    j^*(r_0)= & G(r_0)+ \log(r_0)\left[ \left( \frac{\phi^\alpha_0}{2}+\frac{\alpha^2}{16}+\frac{D(r_0)}{2}(1-\log(r_0))\right) \left(g^*(r_0)*i^*(r_0)+h^*(r_0)\right) \right]\\
    &+ \left(\frac{\phi^\beta_0}{2}+\frac{\alpha^2}{16} \right)i^*(r_0)D(r_0)\log(r_0),
\end{split}
\end{equation*}
\begin{equation*}
  F(r_0)=-4C(r_0)+2C(r_0)\frac{a^2 \log \left( \frac{a}{r_0}\right)}{r_0^2-a^2},
\end{equation*}
\begin{equation*}
\begin{split}
  G(r_0)= &\frac{\phi_0^\alpha}{8} D(r_{0})\log(r_0)+\frac{1}{2}\left(\phi_0^\alpha \gamma^2 +\frac{1}{4}\right)\left(D(r_0) \log(r_0) \right)+\log(r_0) \left( \frac{D(r_0)}{96 \gamma^2}-\frac{(D(r_0))^2}{16}+\frac{1}{8}(D(r_0))^2 \log(r_0)\right)\\
  &+\log(r_0) \left(D(r_0)\left(\phi_0^\alpha \gamma^2 +\frac{1}{4}\right) \left(\frac{1}{16 \gamma^2}+\frac{D(r_0)}{2}(1-\log(r_0)) \right)+\frac{D(r_0)}{16}\left( \phi_0^\beta +D(r_0)(2-\log(r_0))\right) \right),
\end{split}
\end{equation*}
\begin{equation*}
  H(r_0)=\frac{C(r_0)}{4}\left(a^2+4a^2\log\left(\frac{a}{r_0}\right)\right),
\end{equation*}
\begin{equation*}
  J(r_0) = -\frac{(r_0^2-1)^2}{32 \gamma^2}-\frac{(1-r_0^2)}{4}\left(D(r_0)(2-\log(r_0))+\frac{1}{4 \gamma^2} \right)+\frac{C(r_0)r_0^2}{4},
\end{equation*}
\begin{equation*}
\begin{split}
  L(r_0)= &\frac{r_0^2-1}{4}-\frac{3r_0^2+\frac{1}{r_0^2}}{8 \phi_{*}^\alpha}+\frac{\gamma^2}{2 \phi_{*}^\alpha}D(r_0)\left( \frac{1}{r_0^2}-1 \right)-\frac{\gamma^2}{\phi_{*}^\alpha}\left( D(r_0)\log(r_0)-\frac{1}{4 \gamma^2}\right)\left( 1+\frac{1}{r_0^2}\right)+\frac{2 \gamma^2}{\phi_{*}^\alpha}\frac{D(r_0)\log(r_0)}{r_0^2}\\
  &+2 \beta^2 C(r_0)+2 \beta^2 C(r_0)\frac{a^2 \log \left( \frac{a}{r_0}\right)}{r_0^2-a^2}\left( 1+\frac{a^2}{r_0^2}\right)+2 \beta^2 C(r_0)\frac{a^2 \log \left( \frac{a}{r_0}\right)}{r_0^2},
 \end{split}
\end{equation*}
\begin{equation*}
\begin{split}
  M(r_0)=&\frac{1}{16 \gamma^2}(1-r_0^2)(1+3r_0^2)+\frac{D(r_0)}{4}(1+3r_0^2)+\frac{1-2r_0^2}{2}\left(D(r_0)\log(r_0) -\frac{1}{\gamma^2}\right)-\frac{D(r_0)}{2}\log(r_0)\\
  &+\frac{\phi_{*}^\alpha \beta^2}{\gamma^2}\left( C(r_0)a^2 \log \left( \frac{a}{r_0}\right)-C(r_0)a^2\log \left( \frac{a}{r_0}\right)+C(r_0)\frac{3r_0^2}{2}\right),
  \end{split}
\end{equation*} 

\end{document}